\documentclass[conference]{IEEEtran}
\IEEEoverridecommandlockouts

\usepackage{tikz}
\usepackage{amsmath}
\usepackage{amssymb}
\usepackage{multirow}
\usepackage{url}
\usepackage{framed}
\usepackage{xcolor}
\usepackage{filecontents}
\usepackage{rotating}
\usepackage{enumitem}
\usepackage{caption}
\usepackage{subcaption}
\usepackage{comment}
\usepackage{float}
\usepackage{algorithmic}
\usepackage[noline,ruled,algo2e,linesnumbered]{algorithm2e}
\usepackage{xspace}
\usepackage[rightcaption]{sidecap}
\usepackage[normalem]{ulem}

\newcommand{\autotuning}{autotuning\xspace}
\newcommand{\revision}[2]{#2}

\definecolor{darkpastelgreen}{rgb}{0.01, 0.75, 0.24}

\iftrue
\newcommand{\SM}[1]{\textcolor{darkpastelgreen}{SM: #1}}
\newcommand{\RE}[1]{\textcolor{orange}{Romain: #1}}
\newcommand{\PB}[1]{\textcolor{red}{Prasanna: #1}}
\newcommand{\JK}[1]{\textcolor{blue}{JK: #1}}
\newcommand{\MD}[1]{\textcolor{green}{MD: #1}}
\else
\newcommand{\SM}[1]{}
\newcommand{\RE}[1]{}
\newcommand{\PB}[1]{}
\newcommand{\JK}[1]{}
\newcommand{\MD}[1]{}
\fi

\newcommand{\cD}{\mathcal{D}} 
\newcommand{\cC}{\mathcal{C}} %
\newcommand{\cI}{\mathcal{I}} %
\newcommand{\cR}{\mathcal{R}} %
\newcommand{\vect}[1]{#1}

\newif\ifanon 

\let\subparagraph\paragraph
\usepackage[compact]{titlesec}

\titlespacing{\section}{0pt}{1ex}{1ex}
\titlespacing{\subsection}{0pt}{1ex}{0.5ex}
\titlespacing{\subsubsection}{0pt}{0.5ex}{0.5ex}

\DeclareMathSizes{9}{8.2}{5}{3}
\begin{document}

\title{HPC Storage Service Autotuning Using \\ Variational-Autoencoder-Guided \\ Asynchronous Bayesian Optimization
}

\ifanon

\author{
    \IEEEauthorblockN{
        Anonymous authors
    }
}

\else

\author{
    \IEEEauthorblockN{
        Matthieu Dorier\IEEEauthorrefmark{1}\textsuperscript{\textsection},
        Romain Egele\IEEEauthorrefmark{1}\IEEEauthorrefmark{2}\textsuperscript{\textsection},
        Prasanna Balaprakash\IEEEauthorrefmark{1},
        Jaehoon Koo\IEEEauthorrefmark{1},\\
        Sandeep Madireddy\IEEEauthorrefmark{1},
        Srinivasan Ramesh\IEEEauthorrefmark{3},
        Allen D. Malony\IEEEauthorrefmark{3},
        and Rob Ross\IEEEauthorrefmark{1}
    }
    \IEEEauthorblockA{\IEEEauthorrefmark{1}Argonne National Laboratory, Lemont, IL
    -- \{mdorier,pbalapra,jkoo,smadireddy,rross\}@anl.gov}
    \IEEEauthorblockA{\IEEEauthorrefmark{2}Universit\'e Paris-Saclay, France
    -- romain.egele@universite-paris-saclay.fr}
    \IEEEauthorblockA{\IEEEauthorrefmark{3}University of Oregon, Eugene, OR
    -- \{sramesh,malony\}@cs.uoregon.edu}
}

\fi

\maketitle

\ifanon
\else
\begingroup\renewcommand\thefootnote{\textsection}
\footnotetext{These authors contributed equally to the work.}
\endgroup
\fi

\begin{abstract}
Distributed data storage services tailored to specific applications have grown popular in the high-performance computing (HPC) community as a way to address I/O and storage challenges. These services offer a variety of specific interfaces, semantics, and data representations. They also expose many tuning parameters, making it difficult for their users to find the best configuration for a given workload and platform.

To address this issue, we develop a novel variational-autoencoder-guided asynchronous Bayesian optimization method to tune HPC storage service parameters. Our approach uses transfer learning to leverage prior tuning results and use a dynamically updated surrogate model to explore the large parameter search space in a systematic way.

We implement our approach within the DeepHyper open-source framework, and apply it to the autotuning of a high-energy physics workflow on Argonne's Theta supercomputer. We show that our transfer-learning approach enables a more than $40\times$ search speedup over random search, compared with a $2.5\times$ to $10\times$ speedup when not using transfer learning. Additionally, we show that our approach is on par with state-of-the-art autotuning frameworks in speed and outperforms them in resource utilization and parallelization capabilities.
\end{abstract}


\section{Introduction}
\label{sec:Introduction}
Distributed data and input/output (I/O) services have become popular in high-performance computing (HPC) to replace traditional parallel file systems~\cite{ross2020mochi}. They range from multiuser, high-speed storage systems such as burst buffers~\cite{liu2012role,sato2014user,ovsyannikov2016scientific}, to transient, application-specific services providing processing capabilities such as in situ analysis~\cite{childs2020terminology,moreland2011examples,huang2020comprehensive}. These systems aim to improve I/O and storage performance by moving away from file-based interfaces and from the POSIX semantics, instead providing specific interfaces and optimizations that can be tailored to individual applications. An example of such a distributed storage service is HEPnOS~\cite{HEPnOS}, an in-memory object store for high-energy physics (HEP) applications developed by Argonne National Laboratory and FermiLab.

%
%
Like parallel file systems, data services can be incredibly complicated to configure and tune. \revision{}{Contrary to parallel file systems, however, they typically live in user-space and provide more ways for the user to configure them for their specific use-case.} They consist of many inter-related software components that handle different aspects of the service (threading, networking, storage, scheduling), each providing a number of parameters that can be tuned for best performance. From the composition of these building blocks, there emerge even more ways to configure the whole service, such as deciding how they share common resources (CPUs, cores, memory, network, storage devices) and how they are deployed on the physical hardware. Applications that use these services also become more difficult to configure, especially as optimizations such as batching, collective I/O, prefetching, and asynchronous I/O come into play and as these applications are chained together to form workflows. A configuration of the storage service that works well for one step of the workflow may perform poorly for the next, or at a different scale, or on another platform. Given the complexity of these data services, manual tuning is cumbersome and time consuming at best and can easily lead to missed opportunities for better configurations. Hence, a critical need exists for tools and methods that automatically tune not just data services but the entire workflows that use them, searching for well-performing configurations in a given context, and doing so while consuming little time and few resources.


Empirical performance tuning, also known as \emph{\autotuning}, is \revision{}{a hot topic in software optimization nowadays, and} a promising approach for HPC storage service tuning. In this approach, the user exposes the tunable parameters and defines the range of values that each parameter can take; a search method is then used to explore the parameter space by executing different parameter configurations on the target platform. The challenge for HPC storage services \autotuning stems from the complexity of the workflow and the search space. First, several tunable parameters can be interdependent, requiring an execution of the complete workflow on the target platform for a given parameter configuration. Consequently, each parameter evaluation can become expensive. Second, the large number of parameters gives rise to a large search space, which requires sophisticated search methods that can find high-performing configurations in a reasonable search time. Third, given the availability of HPC resources, search methods should leverage them to scale and reduce search time and improve solution quality. Fourth, HPC storage service tuning is not a one-time campaign. Due to changes in workloads, software, and platforms, one needs to run \autotuning regularly. While \autotuning as a whole is a computationally expensive process, the similarity in the \autotuning tasks presents opportunities for leveraging the knowledge gained from one \autotuning campaign to the next. Examples include (1) the user deciding to increase the budget for tuning, but seeking to use the results from a number of smaller \autotuning runs that were performed previously; (2) the user seeking to leverage \autotuning results from a small scale to speed up the \autotuning for large-scale runs; or (3) the user introducing new parameters for \autotuning but seeking to reuse the results for old parameters from previous runs.

\revision{}{
From a mathematical optimization perspective, the \autotuning problem can be formulated as the mixed integer nonlinear optimization problem with computationally expensive black-box objective function, one of the hardest optimization problems to solve. Bayesian optimization is one of the promising approaches for solving these classes of problems ~\cite{shahriari2015taking,bischl2017mlrmbo,bartz2016survey}. Typically, Bayesian optimization is applied in a sequential setting, where the search proposes one parameter configuration for evaluation at each iteration. Distributed Bayesian optimization methods leverage HPC resources to perform simultaneous parameter evaluations to find high-performing configurations in a short wall clock-time. Based on the way in which the simultaneous parameter evaluations are performed,  distributed Bayesian optimization can be grouped into batch synchronous and asynchronous methods. In the former, the search selects a batch of parameter configurations and waits until the evaluations are completed before proceeding to the next iteration. However, this approach results in the waste of HPC resources when the evaluation times are different or diverse. Specifically, the HPC nodes that completed the evaluations faster have to wait until every other evaluations have completed. The asynchronous method overcomes this issue; here as soon as an evaluation is completed, the search method uses the evalaution  result and suggests a new parameter configuration for evaluation. In our setting, given that we are optimizing the run time, the parameter configurations that complete the evaluation faster will update the model frequently and thus increase the chance of sampling more high-performing configurations.    
}


In this paper we develop and apply a new transfer-learning-based search method to tune the parameters of HPC storage services. Our approach adopts a distributed, asynchronous Bayesian optimization method that (1) transfers the results from related \autotuning runs; (2) relies on a dynamically updated and computationally cheap surrogate model to learn the relationship between input configurations and the observed performance; (3) uses the surrogate model to navigate the search space by simultaneously evaluating multiple input configurations; and (4) finds high-performing configurations by evaluating fewer input configurations on the platform. \revision{}{Frameworks such as DeepHyper~\cite{balaprakash2018deephyper}, which we rely on in this work, provide distributed asynchronous BO capabilities. The key novelty of our proposed method consists in \emph{the way it performs systematic transfer learning that leverages the results from similar or related tuning runs}}. This is achieved by using the variational autoencoder, a neural-network-based generative modeling approach, to capture the cross correlations of the high performing configurations of the prior autotuning runs. The joint probability distribution of the high performing configurations is then used to guide the search within Bayesian optimization. This enables the search to focus on promising regions of the search space from the start of the search and signficantly reduces the time required for finding high-performing configurations on new related autotuning runs. Our key contributions in this paper are the following. 
\begin{itemize}[leftmargin=*]
    \item We propose \emph{variational-autoencoder-guided asynchronous Bayesian optimization} (VAE-ABO), a new approach for tuning the parameters of HPC storage services.  We implement VAE-ABO within the DeepHyper framework and make our approach available as an open source software.
    \item We demonstrate the efficacy of the proposed method in a number of settings. Using a high-energy physics workflow, we show that transfer-learning speeds up the search for high-performing configurations and enables faster convergence towards the best configurations, increased resource utilization, and increased number of evaluations.
\end{itemize}

\section{Motivating use-case: HEPnOS and the NOvA event-selection workflow}
\label{sec:Motivation}
Since 2018, the SciDAC ``HEP-on-HPC''~\cite{HEPonHPC} project has brought together physicists and computer scientists from a number of institutions to solve the challenges of leveraging HPC resources to solve large-scale HEP problems. One outcome of this project was HEPnOS, a storage service designed by Argonne National Laboratory to store and provide access to billions of ``event'' data produced by particle accelerators. HEPnOS serves as a centerpiece for a number of HEP analytic workflows, one of which is the event-selection workflow~\cite{paterno2019parallel} for the NOvA experiment~\cite{NOvApaper, NOvA}.

As a highly configurable data service, HEPnOS can be finely tuned for specific use cases. However, finding an optimal configuration for a particular application on a particular platform remains a challenge. Researchers at Argonne and Fermilab have spent the past few years trying to understand how each parameter influences performance, relying on trial and error, manual tuning, benchmarking, and profiling. In this paper we aim to automatize this tuning process by using parameter-space exploration and Bayesian optimization. 

This section dives into some technical details of HEPnOS and the event-selection workflow relevant to our endeavor. 

\subsection{HEPnOS storage system}
\label{sec:Motivation:hepnos}

HEPnOS uses components from the Mochi project~\cite{ross2020mochi}. Thus it relies on Mercury~\cite{soumagne2013mercury} for remote procedure calls and remote direct memory access (RDMA), and on Argobots~\cite{seo2017argobots} for thread management. The Margo~\cite{Margo} library binds Mercury and Argobots together to provide a simple abstraction for developing remotely accessible microservices.

On top of Margo, HEPnOS uses the Yokan microservice~\cite{Yokan} to provide key/value storage capabilities and the Bedrock microservice~\cite{Bedrock} to provide bootstrapping and configuration capabilities.

HEPnOS stores HEP data in the form of a hierarchy of \emph{datasets}, \emph{runs}, \emph{subruns}, \emph{events}, and \emph{products}, the products carrying most of the payload in the form of serialized C++ objects. These constructs are mapped onto a flat key/value namespace in a distributed set of Yokan databases instances. For more technical details on how this mapping is done, we refer the reader to HEPnOS' extensive online documentation.\footnote{\url{https://hepnos.readthedocs.io}}

Of interest to the present work is the fact that some configuration parameters of HEPnOS are critical to its performance, including its number of database instances, how these databases map to threads, how it schedules its operations, down to low-level decisions such as whether to use blocking \texttt{epoll} or busy spinning for network progress.

Thanks to the Mochi Bedrock component, which provides configuration and bootstrapping capabilities to Mochi services, all these parameters can easily be provided from a single JSON file that describes which components form the service and how they should be configured. This extensive configurability is critical to the work presented in this paper, and is what distinguishes a storage service such as HEPnOS from more traditional storage systems such as a parallel file system.

\subsection{NOvA event-selection workflow}
\label{sec:Motivation:workflow}


\begin{figure*}
    \centering
    \includegraphics[width=\textwidth]{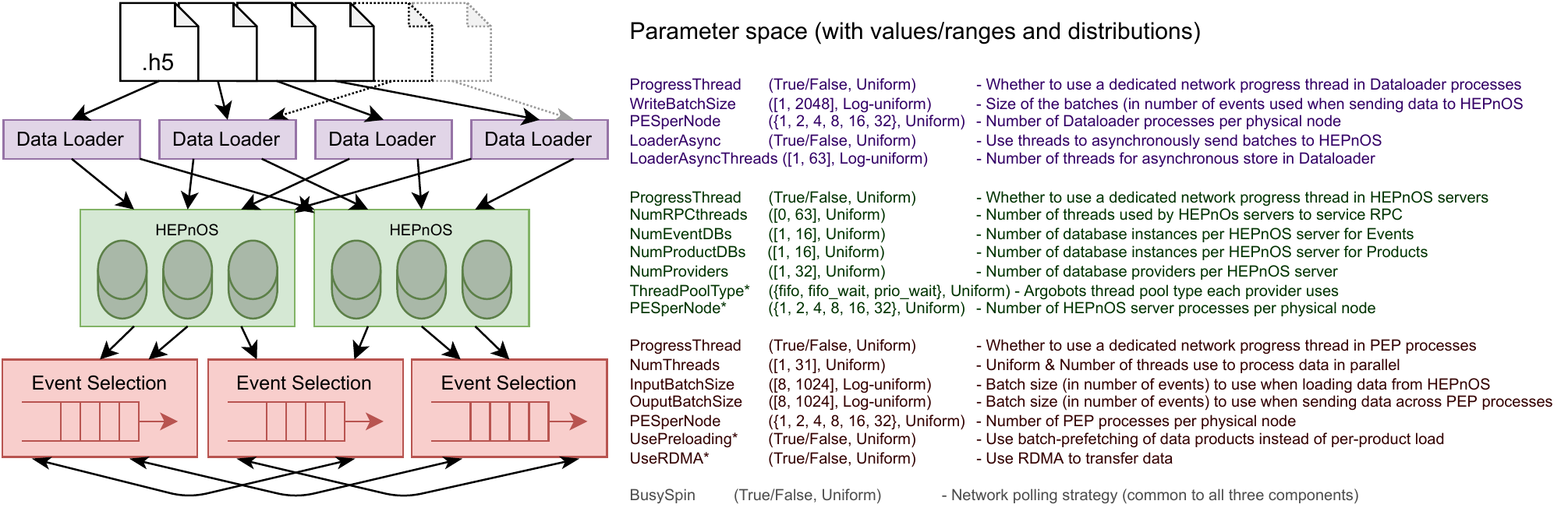}
    \caption{\revision{}{Overview of the HEP event-selection workflow. A set of HDF5 files containing tables of event data is read by a parallel application, the data loader, which converts them into arrays of C++ objects that are then stored in HEPnOS as \emph{products} associated with \emph{events}. In the event-selection step, all the events contained in a given \emph{dataset} are read and processed in parallel to search for events matching specific criteria. Each component of this workflow presents a set of configurable parameters. The parameters considered in this work are shown on the right, with colors matching their associated component. Parameters with a * are used only in the extended search space (see Section~\ref{sec:Evaluation:methodology:setup}).}}
    \label{fig:Motivation:schemas}
\end{figure*}

\revision{}{As shown in Figure~\ref{fig:Motivation:schemas},}
HEPnOS is used as a distributed, in-memory storage system for HEP workflows. In this work we focus on the NOvA event-selection workflow, which consists of two steps: data loading, and parallel event processing.
A set of HDF5 files containing tables of event data is read by a parallel application, the data loader, which converts them into arrays of C++ objects that are then stored in HEPnOS as \emph{products} associated with \emph{events}. In the event-selection step, all the events contained in a given \emph{dataset} are read and processed in parallel to search for events matching specific criteria.

%
%
\subsubsection{Data loading}
\label{sec:Motivation:workflow:dataloader}

In practice, the event selection workflow does not operate directly on data as it is produced by the particle accelerator. The raw data is first stored into files, either in HDF5~\cite{hdf5} or in ROOT~\cite{ROOT} format. While these files can be shared across institutions easily, they produce an I/O bottleneck when it comes to reading them from a large number of processes. Hence in HEPnOS-based workflows they need to be loaded into HEPnOS prior to their data being processed.

The dataloader is in charge of this task. It is a parallel, MPI-based application that takes a list of HDF5 files, converts them into C++ objects, and stores them into HEPnOS. Since the amount of data differs across HDF5 files, the dataloader does not distribute the work in a static manner across its processes. Instead, a list of files is maintained in one process, and all the processes pull work from this shared list of files until all the files have been loaded.

Several optimizations are available in the dataloader, including batching of events and products (the mapping of events and products to HDF5 files on the one hand and to databases in HEPnOS on the other hand is such that all the events coming from the same file will end up in the same database, and similarly for products) and overlapping the loading of a file with the storage of data from a previous file into HEPnOS. These optimizations can be turned on or off and configured in various ways. Along with job-related parameters (number processes, number of threads, mapping to CPUs), the dataloader offers many configuration parameters that can be tuned to achieve good performance.

%
%
\subsubsection{Parallel event processing}
\label{sec:Motivation:workflow:pep}

The second step of the workflow, \emph{parallel event processing} (PEP), consists of reading the events and some products associated with them, and performing some computation on the data to determine events of interest. If events are stored in $N$ databases in HEPnOS, $N$ processes of the PEP application will list them, each accessing one database.
They will end up filling a local list of events (\texttt{<dataset id, run number, subrun number, event number>} tuples). All the processes pull events from either their local queue or by requesting batches from other processes. Each event is processed first by loading the data products associated with it, then by performing computation on these products.

The PEP application provides a benchmark of its I/O part (loading events and products) that simulates computation. We use this benchmark in place of the real PEP application in this paper since we are interested in autotuning only the I/O aspects of this workflow.

Just like HEPnOS and the data loader, optimizations are in place to improve I/O performance in the PEP application and benchmark: look-ahead prefetching when reading from HEPnOS, batching of events when they are loaded and when they are sent from one process to another, batching of data products, and multithreaded processing of events inside each process. All these optimizations come with their own set of tunable parameters that can influence the overall performance of the workflow.

%
%
\subsection{Challenges of (auto)tuning the workflow}

While the parameters of a single workflow component cannot be tuned independently from one another, they also cannot be tuned independently from the parameters of other components. As an example, what could seem like an optimal number of threads in the PEP application could in turn influence the optimal number of databases in HEPnOS, which could influence the batch size used by the dataloader when storing events into HEPnOS. Manually tuning such a workflow becomes rapidly intractable, in particular as new optimizations (hence new parameters) are implemented, new steps are added to the workflow when the workflow scales up, or when it is ported to a new platform. This situation motivated us to investigate ways of automatically tuning such a workflow using parameter space exploration and machine learning.

Parameter-space exploration enables defining the list of tunable parameters and their domain. It then uses a ``meta-workflow'' to execute multiple instances of the workflow in parallel, with different configurations, in order to measure how fast they complete when given a relatively small number of HDF5 files as input. This method, however, poses an important challenge: evaluating many configurations consumes a lot of computation resources, especially as the parameter space grows and as the use case gets more complex.
Hence, this paper focuses on ways to speed up such an exploration process by (1) using asynchronous Bayesian optimization to model the workflow performance as a function of its configuration and (2) using transfer learning to go from a small-scale, one-step workflow with few parameters to a large-scale, full workflow with more parameters. The next section presents these two contributions.

\section{Proposed approach: Transfer-learning-enabled autotuning}
\label{sec:Contribution}

We now formally define the class of \autotuning problems we consider. We assume that we are given a \autotuning problem with a single performance metric and a set of feasible tuning parameters, and pose this as the following optimization problem:
\begin{equation}
\max_x \left\{ f(x) : x=(x_\cI,x_\cR,x_\cC) \in \cD \right\}.
 \label{eq:genprob}
\end{equation}
We denote the $n$ tunable parameters by a vector $x=(x_\cI,x_\cR,x_\cC)$, where $x$ can be partitioned into three types of parameters  $\cI$, $\cR$, and $\cC$, respectively denoting integer parameters with a natural ordering, continuous parameters that take real values, and categorical parameters to explicitly model parameters whose values have no special ordering. Typically, the feasible set $\cD$ is defined by a set of constraints on $x$. 


The problem given in Eq.~\ref{eq:genprob} is a nonconvex mixed-integer nonlinear program (MINLP), one of the most challenging problems in optimization. Moreover, the derivatives of relaxations of $f$ with respect to the decision variables, $\nabla_x f$ cannot be obtained, and the optimization needs to be carried out by evaluating $f(x)$. This makes Eq.~\ref{eq:genprob} a black-box MINLP, adding further complexity.

\subsection{Asynchronous Bayesian optimization}
\label{abo}

Bayesian optimization (BO) is a promising approach for tackling the class of autotuning problems described in Eq.~\ref{eq:genprob}. BO relies on fitting a dynamically updated surrogate model that tries to learn the relationship between $x$ (input) and $f(x)$ (output) by modelling $p(y|x)$, which quantify the stochastic/noise aspects of the evaluated workflow. The key feature of the surrogate model is the uncertainty quantification capability. For any evaluated $x$, the model returns the predictive mean and variance. The latter is leveraged to assess how uncertain the surrogate model is in predicting $f(x)$. The surrogate model is cheaper to evaluate than the objective function evaluation and can be used to identify promising regions, where the surrogate model is then iteratively refined by selecting new inputs that are predicted by the model to be high-performing (exploitation of search space) or that can potentially improve the quality of the surrogate model (exploration of search space). BO navigates the search space by switching between exploration and exploitation to find high-performing configurations. 

\revision{}{
We focus on \emph{sampling-based} BO that selects an input for evaluation as follows:  A large number of unevaluated configurations are sampled from the feasible set $\cD$. The BO uses a dynamically updated surrogate model $M$ to predict a point estimate (mean value) $\mu(x_i)$ and standard deviation $\sigma(x_i)$ for each sampled configuration $x_i$. The sampled configurations are then ranked by using the lower-confidence bound (LCB) acquisition function~\cite{shahriari2015taking} given by
\begin{equation}
    LCB(x_i) = \mu(x_i) - \kappa \cdot \sigma(x_i),
    \label{eqn:ucb}
\end{equation}
where $\kappa \geq 0$ is a parameter that controls the trade-off between exploration and exploitation. When $\kappa$ is set to zero, the search performs only exploitation (greedy); a large value for $\kappa$ results in pure exploration. A balance between exploration and exploitation is achieved when $\kappa$ is set to an appropriate value such as 1.96 for a 95\% confidence interval around the mean estimate. 

While classical BO methods are designed to perform sequential evaluations of input configurations, asynchronous BO takes advantage of parallel computing resources to perform simultaneous evaluations of parameter configurations. The asynchronous BO that we adopt follows a manager-workers distributed architecture. To generate multiple configurations at the same time, BO leverages a multipoint acquisition function based on a constant liar strategy~\cite{hiot_kriging_2010}. This approach starts by selecting a parameter configuration that minimizes the LCB function. Then a copy of the model $M'$ is updated with the selected configuration and a dummy value (lie) corresponding to the minimum value of collected objectives (performance). This encourages sampling other configurations from nearby regions. The next parameter configuration is obtained by minimizing the LCB function using the updated model. This process is repeated until the required number of configurations is sampled (i.e., as many as the number of idle workers). While numerous advanced asynchronous BO techniques are available, the constant liar strategy was chosen because of its  computational simplicity and minimal overhead. In this way, the asynchronous BO will have the capability to create a number of configurations for simultaneous evaluation within a short period of time, thus significantly reducing idle workers.
}

\subsection{Variational-autoencoder-guided Bayesian optimization for transfer learning}



\begin{algorithm2e}[!ht]
\small
\DontPrintSemicolon
\SetInd{0.25em}{0.5em}
\SetAlgoLined
\SetKwInOut{Input}{inputs}\SetKwInOut{Output}{output}
\SetKwFunction{RandomPoint}{random\_point}
\SetKwFunction{SubmitEval}{submit\_evaluation}
\SetKwFunction{GetFinishedEval}{get\_finished\_evaluations}
\SetKwFunction{Push}{push}
\SetKwFunction{EmptyList}{EmptyList}
\SetKwFunction{RandomSample}{random\_sample}
\SetKwFunction{SelectParent}{select\_parent}
\SetKwFunction{Mutate}{mutate}
\SetKwFunction{Tell}{tell}
\SetKwFunction{Ask}{ask}

\SetKwFor{For}{for}{do}{end}

\Input{$\mathbf{H}_{p}$: search history from previous autotuning, $q\%$: quantile value for high-performing parameter configurations selection,
$\cD^p$: previous parameter space, $\cD^c$: current parameter space, $W$: workers}
\Output{$x^{curr*}$: best configuration from the current autotuning, $y^{curr*}$, the performance metric of the $x^{curr*}$, $\mathcal{H}$, evaluations from the search \\}
    {\color{orange} \tcc{Informative prior initialization}}
    
    {$\mathbf{Q}_{p}$ $\leftarrow$ \texttt{subset}($\mathbf{H}_{p}$,$q\%$)}\\
    
    {\color{orange} \tcc{Fit tabular variational autoencoder using Bayesian Optimizer}}
   
    {$\mathcal{P}$ $\leftarrow$  \texttt{TVAE}($\mathbf{Q}_{p}$)}\\
    
    {\color{orange} \tcc{User-defined prior initialization for new parameters}}
    \ForEach{$x_j \in \cD^c$}{
        \uIf{$x_j \not\in \cD^p$}{
            \uIf{$x_j \in \mathcal{I}\ or\ \mathcal{R}$ }{
               $\mathcal{P}(x_j) = \texttt{Uniform}(l_j,u_j)$ 
               }
            \Else{
                $\mathcal{P}(x_j) = \texttt{Multinoulli}(p_j)$
            }
        }
    }
    
    $\mathcal{H} \leftarrow \{\}$\\
    {$optimizer$ $\leftarrow$ \texttt{Bayesian\_{Optimizer}}($\mathcal{D}^c$,$\mathcal{P}$)}\\
    
    {\color{orange} \tcc{Initialization of BO}}
    \For{$i\leftarrow 1$ \KwTo $W$}{
        $ x_i \leftarrow$ \texttt{sample\_configuration}{$(\cD^c$ | $\mathcal{P})$}\\
        \SubmitEval{$x_i$} \tcp{Nonblocking}
    }
    
    {\color{orange} \tcc{Optimization loop of BO}}
    \While{stopping criterion not met}{
        \tcp{Query results }
        ($\mathcal{X}_e, \mathcal{Y}_e$) $\leftarrow$ \GetFinishedEval()\\
        $\mathcal{H} \leftarrow \mathcal{H} \cup  (\mathcal{X}_e, \mathcal{Y}_e)$\\
        {\tcp{Generate parameter configs}}
        { $optimizer.$\Tell{$\mathcal{X}_e, \mathcal{Y}_e$}\\
        $\mathcal{X}_{next} \leftarrow$ $optimizer.$\Ask{$|\mathcal{Y}_i|$ | $\mathcal{P}$}}\\
        \SubmitEval{$\mathcal{X}_{next}$} \tcp{Nonblocking}
    }
    $x^{curr*}$, $y^{curr*}$ $\leftarrow$ \texttt{find\_best}($\mathcal{H}$)\\
    return $x^{curr*}$, $y^{curr*}$, $\mathcal{H}$\\
\caption{Variational-Autoencoder-Guided Asyncronous BO (VAE-ABO)}
\label{alg:PGBO}
\end{algorithm2e}

Transfer learning (TL) in \autotuning seeks to transfer the information gained from a previous related search to a new one, to improve either the search efficiency or accuracy, or both. Our TL approach is based on the intuition that the high-performing configurations and their neighborhood from the previous \autotuning run can provide   the parameter subspace of potentially high-performing configurations for a related new \autotuning run. Typically, BO starts with user-defined prior distributions (typically uniform) for each parameter. This process entails that each value for a given parameter has an equal probability of being sampled in the sampling process of BO, specifically, in the initialization phase for bootstrapping the model $M$ and the candidate configuration selection phase to generate a large number of unevaluated configurations. In our TL strategy, we use the high-performing configurations to define informative prior distributions for the parameters.  This is achieved by a generative modeling approach to model the joint probability distribution of the high-performing configurations from the previous related autotuning run. Specifically, we use a variational autoencoder~\cite{7796926} to learn the joint probability distribution and use the trained model for sampling the configurations in the initialization phase and generating unevaluated configurations to select candidates for evaluation in the iterative phase.
By doing so, the sampled configurations in the BO are biased toward the high-performing configurations from the previous run. Consequently, the BO's exploration and exploitation using the surrogate model $M$ take place in the biased regions of the parameter space. We call our approach variational-autoencoder-guided asynchronous BO (VAE-ABO).

Algorithm~\ref{alg:PGBO} shows the pseudo code of our proposed VAE-ABO. It takes the following inputs: $\mathbf{H}_{p}$, the search history from the previous autotuning run comprising the input parameter configurations and their corresponding performance metrics;  $q\%$ the quantile value that defines the high-performing configurations in $\mathbf{H}_{p}$; $s$, the number of samples for VAE evaluation; the parameter domain $\cD^p$ of the previous autotuning run comprising a list of parameters and their types; and parameter domain for the current autotuning run $\cD^c$ with the list of parameters and their types. 

The key component of VAE-ABO is the definition of informative priors using VAE.
It starts by computing the $q\%$ quantile value of the output performance metric in $\mathbf{H}_p$ and selecting a set of high-performing input configurations $\mathbf{Q}_p$ that have their output performance metric greater than or equal to the computed quantile value. The next step is to model the joint probability distribution of the selected high-performing parameter configurations $\mathbf{Q}_p$ using VAE. 
While VAEs are typically used to model complex probability distributions and generate  high-quality realistic samples, they are typically not suitable for our setting because of the tabular nature of the parameter configurations. 
The key modeling challenges of tabular data can be attributed to mixed integer parameter space, multimodality, and imbalanced parameter values. Recently, a new VAE for tabular data called TVAE to model the probability distribution of tabular data has been proposed \cite{xu2019modeling}. 
TVAE is defined by three components: encoder, decoder, and a loss function.  The encoder $q_{\vartheta_{e}} (z \mid \vect{x})$, a multilayered deep neural network, parameterized by $\vartheta_e$, will learn to take the high-performing input configurations  $\vect{x}$ from $\mathbf{Q_p}$ and project them into a lower dimensional stochastic latent space $z$, described by a multivariate Gaussian distribution. 
The latent layer is also referred to as a bottleneck layer because the encoder is tasked with learning an efficient projection of the input data into $z$. 
Let the neural network that is used to approximate the encoder be denoted as $f(.;\vartheta_{e}).$ The encoding is therefore given as
$$z = q_{\vartheta_{e}} (z \mid \vect{x}).$$
The decoder $q_{\vartheta_d} (z \mid \vect{x})$, parameterized by $\vartheta_d$, is another neural network that takes as input samples from the latent space representation of $\vect{x}$ in $z$ and learns to reconstruct $x$. 
The output of the decoder is then provided by applying a neural network function $g(.; \vartheta_d)$ on samples from the encoder.
$$ \hat{\vect{x}} = g(z; \vartheta_d),$$
where $\hat{\vect{x}}$ is the reconstruction of the input $\vect{x}.$ 
The TVAE training involves adapting the neural network parameters $\vartheta_e$ and $\vartheta_d$ to minimize the evidence lower-bound loss using stochastic gradient descent type optimization.

For each parameter $x_j$ in $\cD^c$ but not in $\cD^p$, VAE-ABO defines uninformative prior  (l. 3-10): For an integer or real parameter, the uninformative prior is defined using the uniform probability (l.~6) distribution with the $l_j$ and $u_j$ as bounds. If $x_j$ is a categorical parameter, the uninformative prior is defined using a \textit{multinoulli} probability distribution (l.~12) but parametrized by $p_j$ =  $\{p^1_j, \ldots, p^K_j\}$, where $\forall p^i_j = p$ (all categories have the same probability $p$).  




The asynchronous BO starts by sampling parameter configurations based on the informative prior distribution $\mathcal{P}$ (l.~13-16). 
These configurations are then submitted for evaluation in an asynchronous fashion. In the optimization loop (l.~17-23), the search first waits for completed evaluations (l.~18). As soon as they are available (denoted by sets of inputs $\mathcal{X}_e$ and their corresponding outputs, $\mathcal{Y}_e$, respectively), they are given to the optimizer object to update the surrogate model (l.~20). This is followed by sampling $|\mathcal{Y}_e|$ number of configurations (l.~21) for further evaluations and uses the same three steps as described in Sec \ref{abo}: generate a large number of configurations, use the surrogate model to predict mean and standard deviation, and rank them using UCB. The key difference is in the sample generation: instead of uniform random sampling to generate configurations, the samples are generated from the informative prior $\mathcal{P}$. This ensures that the majority of the sampled configurations are close to $\mathbf{Q}_p$ and consequently, even the exploration part of the search is biased towards configurations that are not far away from $\mathbf{Q}_p$. The loop continues until a user-defined stopping criterion is met (typically wall clock time or maximum number of evaluations). Given the search history $\mathcal{H}$, the maximal performance $y^{curr*}$ and its corresponding parameter configurations are extracted. Finally, $x^{curr*}$, $y^{curr*}$ and $\mathcal{H}$ are returned.

\subsection{Implementation}

\begin{figure}
    \centering
    \includegraphics[width=\columnwidth]{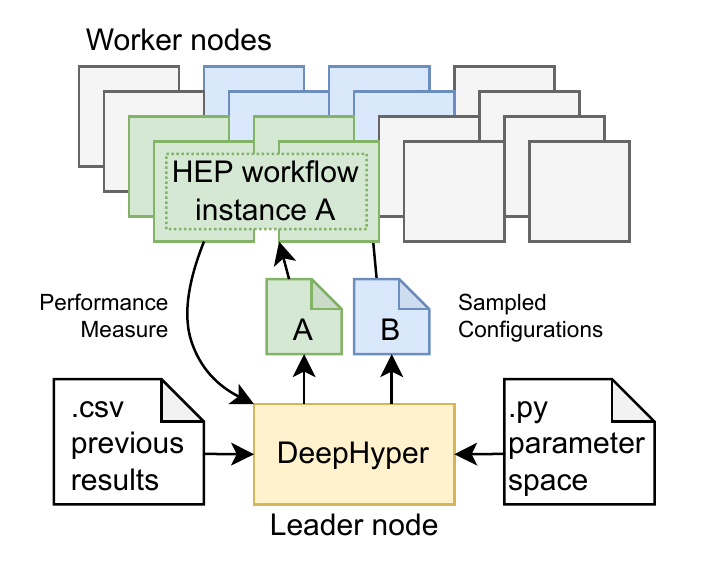}
    \caption{\revision{}{Overview of the DeepHyper-based autotuning framework. DeepHyper takes as input a python script defining the parameter space, and optionally a CSV file containing data from previous experiments for transfer-learning. It asynchronously submits instances of the HEP workflow on worker nodes in parallel, monitors their performance, and updates its internal model to pick the next configurations to evaluate.}}
    \label{fig:Contribution:deephyper}
\end{figure}

We implemented VAE-ABO within the asynchronous BO method of DeepHyper \cite{balaprakash2018deephyper}, a scalable open source Python package that was originally developed for hyperparameter optimization of machine learning methods. \revision{}{For the VAE, we used the same neural network architecure described in ~\cite{7796926}, which is available as a open source package.} The surrogate model used is a random forest regressor. Figure~\ref{fig:Contribution:deephyper} gives an overview of our autotuning framework.

Since DeepHyper aims to maximize a provided objective function, and our aim is to minimize the HEP workflow run time, we define our objective function as
$ - \text{log}(runtime)$. The use of a logarithm allows the search to better discriminate between small values of the run time.

\section{Evaluation}
\label{sec:Evaluation}

In this section we evaluate the effectiveness of our VAE-ABO autotuning approach by using it to search for high-performing configurations of the HEP workflow presented in Section~\ref{sec:Motivation}.

\subsection{Objectives and methodology}
\label{sec:Evaluation:methodology}

Our goal in this evaluation is twofold. First, we want to show that \emph{using transfer learning with VAE-ABO is more effective than performing a new search without prior knowledge}. Second, we want to show that \emph{our VAE-ABO-enabled DeepHyper-based framework is more effective than other state-of-the-art frameworks and methods}.

\subsubsection{Effectiveness metrics}
\label{sec:Evaluation:methodology:metrics}

To achieve our goal, our analysis focuses on five metrics.

\begin{description}[leftmargin=0pt]

\item[Best-performing configuration.] We measure the run time $R_{best}$ of the HEP workflow using the best-performing configuration found by a given approach within a given search time $t_{max}$ (typically 1 hour).

\item[Mean best-performing configuration.] This metric averages the run time of the best-performing configuration known after a search time $t$ for $t \in [0, t_{max}]$. More formally,
\[ E[R] = \frac{1}{t_{max}} \int_{t=0}^{t_{max}} R(t) dt, \]
where $R(t)$ is the run time of the best-performing configuration found after a search time of $t$. It corresponds to the expected best run-time when stopping the search after a time $t$ drawn uniformly between 0 and $t_{max}$. If two approaches reach the same $R_{best}$, this metric allows comparing which one was the fastest converging toward it.

\item[Number of evaluations.] We count the number of instances of the HEP workflow that the approach was able to run within a given search time. Two main factors affect this number: how fast an approach decides on the next configurations to evaluate and how fast these evaluations complete. The latter correlates with how effective a method is at quickly finding and focusing on high-performing configurations.

\item[Worker utilization.] This metric represents the percentage of time the workers spent actually running instances of the HEP workflow. If a method/framework is too slow when deciding which configuration to evaluate next, the workers will stay idle, wasting valuable compute resources. Note that a method can have a high worker utilization but a low number of evaluations (typically if evaluations take more time because the approach picks poorly performing ones).

\item[Search speedup.] This metric measures how much faster a method is at finding good configurations compared with a random sampling. Let $R_{best}^{rand}$ be the run time of the best configuration found by a random sampling after a search time of $t_{max}$. The search speedup of a method is defined by the minimum time it takes for the method to reach this run time, relative to the maximum search time, or more formally
\[ S = \frac{t_{max}}{ \underset{t}{\mathrm{argmin}}(R(t) < R_{best}^{rand})}.\]
When considering multiple repetitions of an experiment, we replace $R_{best}^{rand})$ and $R(t)$ with averages over these repetitions.

\end{description}

\subsubsection{Workflow setup and experiments}
\label{sec:Evaluation:methodology:setup}

We worked with the workflow users to select 20 parameters affecting the performance of the HEP workflow presented in Section~\ref{sec:Motivation}. \revision{Seven of these parameters concern HEPnOS itself; Five concern the data loader; seven concern the event-selection application; and one is common to all components. They range from low-level parameters affecting networking (e.g., polling mechanism) and thread scheduling (e.g., Argobots pool type) to high-level parameters influencing various optimizations (e.g., batch sizes, use of RDMA, concurrency level) and process/thread placement. Three of these parameters are ordinal, eight are categorical, and nine are range-based.}{These parameters are listed in Figure~\ref{fig:Motivation:schemas}.} This parameter space contains around $1.5\times 10^{23}$ distinct configurations.

We consider five setups of the HEP workflow, summarized by the nomenclature hereafter \revision{}{(the first number represents the number of nodes used by each HEP workflow, the second number is the number of steps considered in the workflow, the third is the number of parameters of the parameter space).}
\begin{description}[leftmargin=0pt]
    \item[\revision{}{4n-1s-11p}:] Each workflow instance uses 4 nodes: 1 for HEPnOS and 3 for the applications using it. Only the data-loading part of the workflow is executed. Furthermore, only 3 of the 5 parameters affecting the data loader are considered.
    The parameter space of this setup consists of 11 parameters.
    \item[\revision{}{4n-2s-16p}:] Compared with the previous setup, we enable the event-selection part of the workflow but use only 5 of its 7 parameters, leading to a parameter space of 16 parameters.
    \item[\revision{}{4n-2s-20p}:] We enable the full set of parameters for each of the applications in the workflow.
    \item[\revision{}{8n-2s-20p}:] We scale the previous setup to 8 nodes per workflow instance: 2 for HEPnOS and 6 for its applications.
    \item[\revision{}{16n-2s-20p}:] We scale the previous setup to 16 nodes per workflow instance: 4 for HEPnOS and 12 for its applications.
\end{description}

The Fermilab team kindly provided us with a sample of 200 HDF5 files (totaling 26.5 GiB), which also limits the scale at which we can run instances of the workflow. We use 50 files for 4-node experiments, 100 files for 8 nodes, and 200 files for 16 nodes (weak scaling).

We always perform TL from one type of setup to the next (e.g., from \revision{}{4n-2s-16p} to \revision{}{4n-2s-20p} but not from \revision{}{4n-2s-16p} to \revision{}{8n-2s-20p}). Going from \revision{}{4n-1s-11p} to \revision{}{4n-2s-16p} represents a scenario where the workflow changes. Not only does the parameter space change accordingly, but we can also expect the best configuration for the first part of the workflow to change. \revision{}{4n-2s-16p} to \revision{}{4n-2s-20p} represents a scenario in which we have done a search and now want to include more parameters in our search space.
Scaling from 4 to 8 and 16 nodes represents a scenario in which the conditions of execution of the workflow change but neither the parameter space nor the workflow itself.

\subsubsection{Real platform and simulated runs}
\label{sec:Evaluation:methodology:platform}

The experiments presented in Sections~\ref{sec:Evaluation:tl} and~\ref{sec:Evaluation:all} were carried out on the Cray XC40 ``Theta'' supercomputer~\cite{Theta}, an 11.7-petaflops machine composed of 4,392 nodes, each featuring a 64-core, 1.3-GHz Intel Xeon Phi 7230 processor, and interconnected via a Cray Aries network with a Dragonfly topology. Each experiment is deployed on 128 nodes that DeepHyper uses as workers to execute a HEP workflow instances.

The experiments presented in Section~\ref{sec:Evaluation:surrogate}, which compare our approach with state-of-the art frameworks, are done by training a surrogate model to simulate the behavior of the real HEP workflow. This allows these experiments to run on a local machine that simulates the execution of the workflow on Theta, leading to more reproducibility of our results.

Initial executions of the HEP workflow on 4 nodes with manually selected parameters led to a run time on the order of 90 seconds for the data loader, and 90 seconds for the event selection. Hence, to prevent a bad configuration from taking too long to complete, which would prevent the evaluation of potentially better configurations, we limit the execution time of each workflow instance to 600 seconds (300 seconds per step). If a workflow step (data loader or event selection) has not completed within its time limit, it is killed, and the evaluation returns a run time of $NaN$ (not a number). Similarly if a workflow instance crashes, a run time of $NaN$ is returned.

\subsubsection{Availability and reproducibility}
\label{sec:Evaluation:methodology:reproducibility}

Our modifications to DeepHyper are already upstream. Our code using DeepHyper to autotune the HEP workflow is available on GitHub,\footnote{\revision{}{\url{https://github.com/hepnos/HEPnOS-Autotuning}}} with instructions on how to build and use it. Unfortunately we did not obtain permission from the Fermilab team to make their HDF5 files public at this time.

All our results from Sections~\ref{sec:Evaluation:tl} and~\ref{sec:Evaluation:all} are, however, available in CSV format along with scripts and instructions to analyze them and reproduce all the figures presented in this paper (and many more).\footnote{\revision{}{\url{https://github.com/hepnos/HEPnOS-Autotuning-analysis}}} In total, 115 CSV files are provided, each corresponding to a 1-hour experiment completed on 128 nodes of the Theta supercomputer, representing 51,830 evaluations of the HEP workflow.

All our experiments from Section~\ref{sec:Evaluation:surrogate} can be executed on a desktop machine and are fully reproducible \revision{}{simply by executing the Jupyter notebook provided in our repository. This notebook also provides an extensive description of the parameter space, which we cannot include in the present paper because of space constraints}. We also provide their corresponding CSV files (70 files), should the reader wish to examine them without reproducing the experiments themselves.

\subsection{Evaluating VAE-ABO in DeepHyper}
\label{sec:Evaluation:tl}

Here, we evaluate the benefit of TL using VAE-ABO. We use DeepHyper to run each of the setups presented in Section~\ref{sec:Evaluation:methodology:setup} for 1 hour on 128 nodes. Each run is repeated 5 times.

\begin{figure*}
    \centering
    
    \begin{subfigure}[b]{0.32\textwidth}
        \centering
        \includegraphics[width=\textwidth]{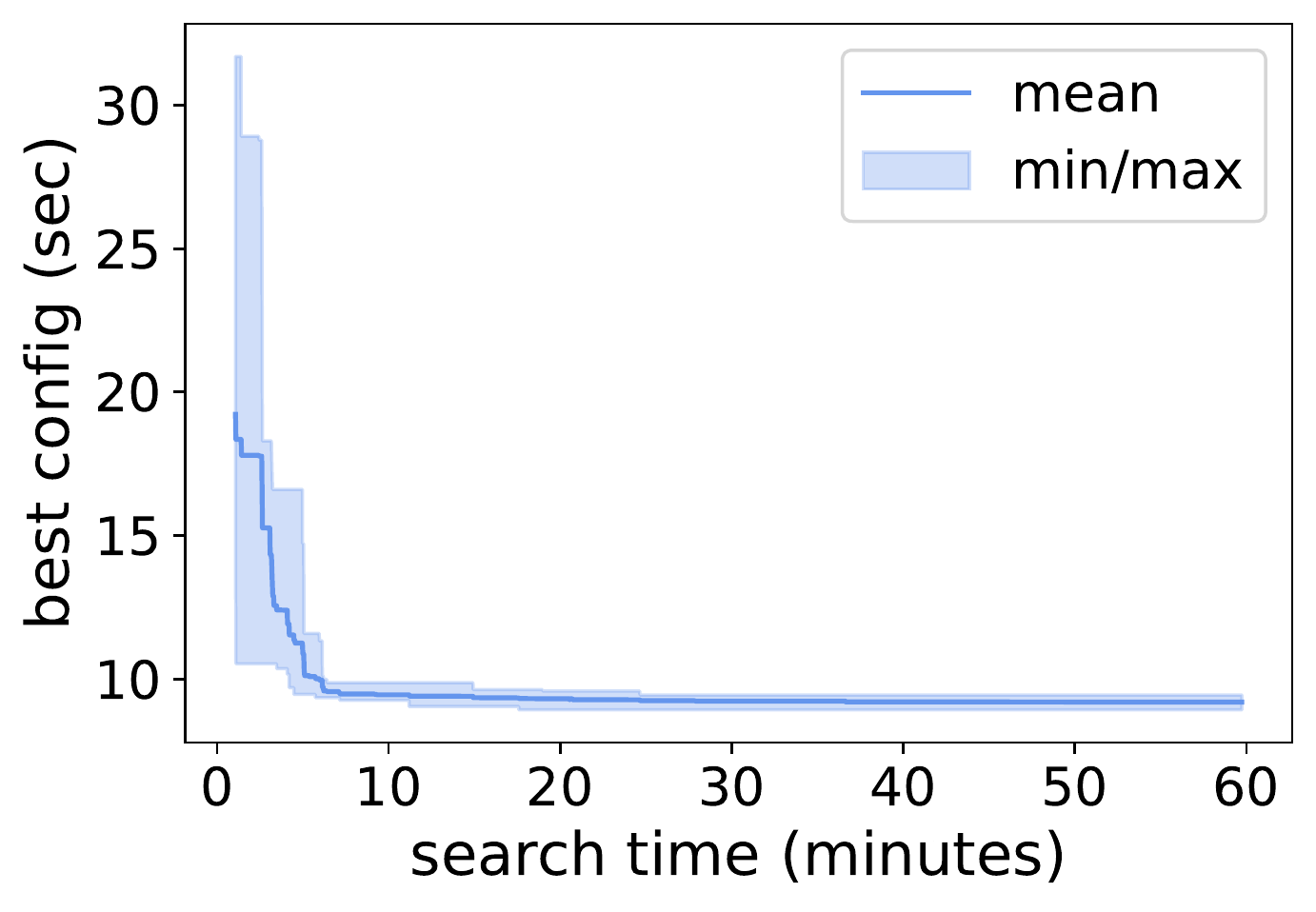}
        \caption{\revision{}{4n-1s-11p}}
        \label{fig:Experiments:tl:4-F-F}
    \end{subfigure}
    \begin{subfigure}[b]{0.32\textwidth}
        \centering
        \includegraphics[width=\textwidth]{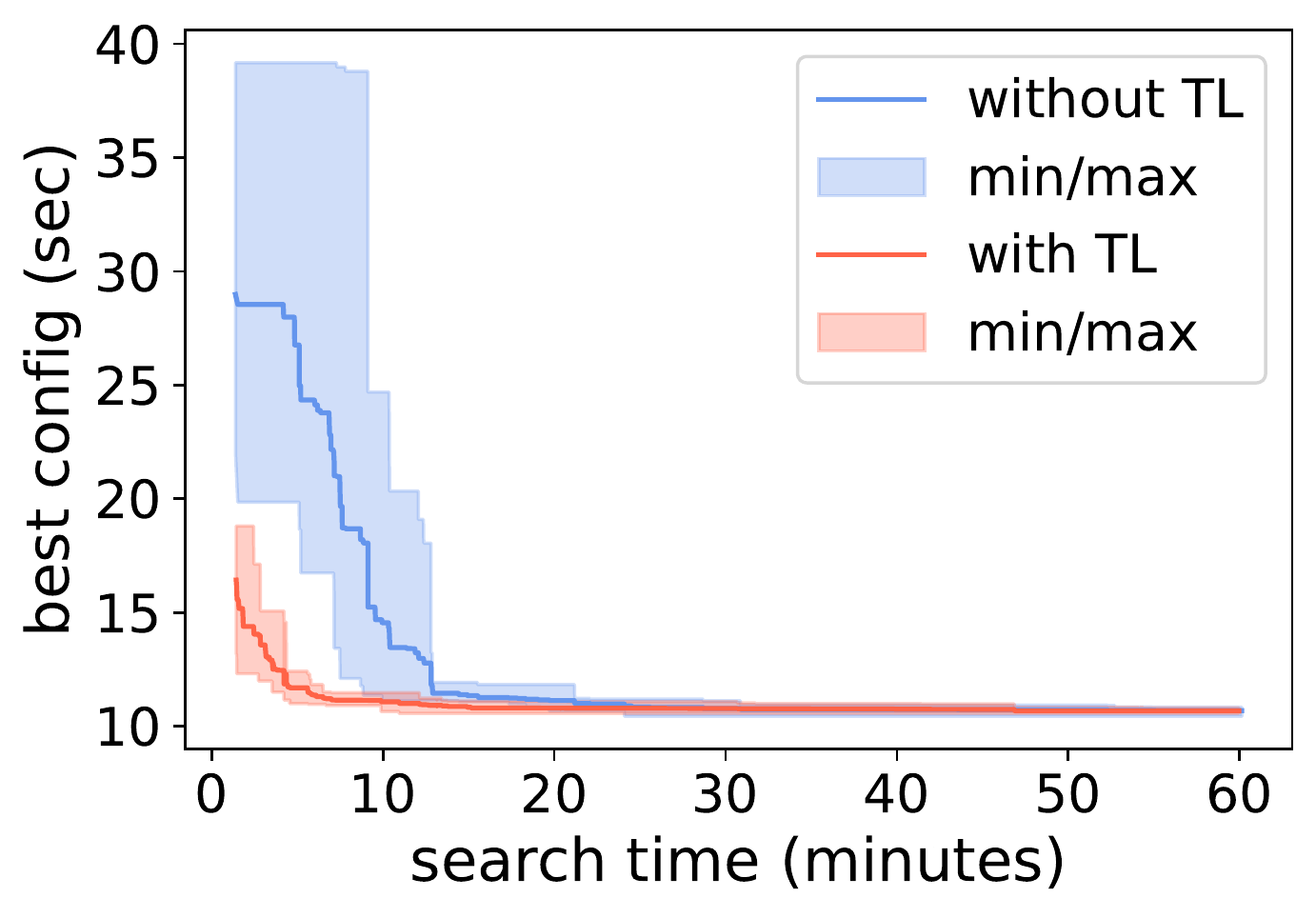}
        \caption{\revision{}{4n-2s-16p}}
        \label{fig:Experiments:tl:4-T-F}
    \end{subfigure}
    \begin{subfigure}[b]{0.32\textwidth}
        \centering
        \includegraphics[width=\textwidth]{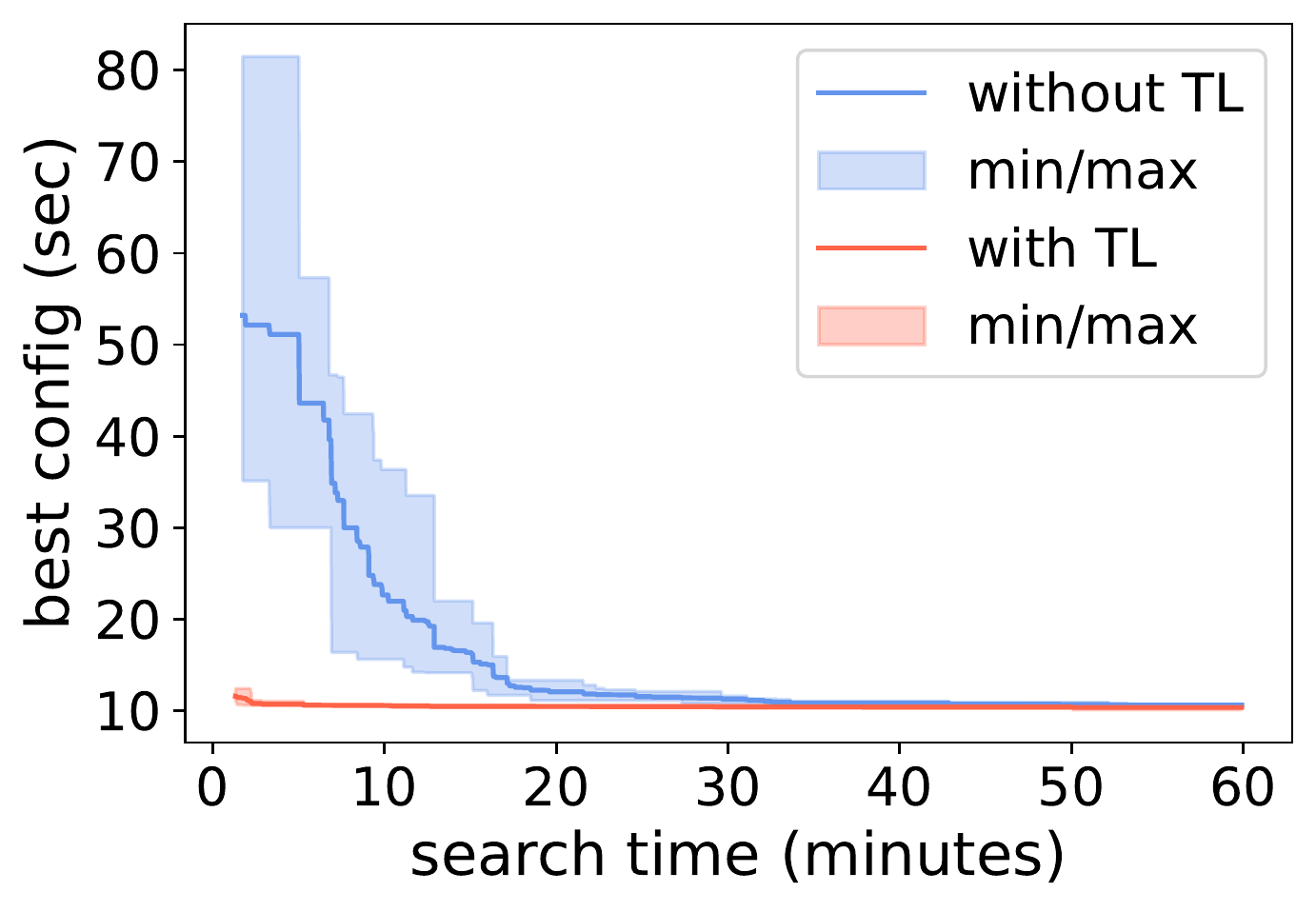}
        \caption{\revision{}{4n-2s-20p}}
        \label{fig:Experiments:tl:4-T-T}
    \end{subfigure}
    \begin{subfigure}[b]{0.32\textwidth}
        \centering
        \includegraphics[width=\textwidth]{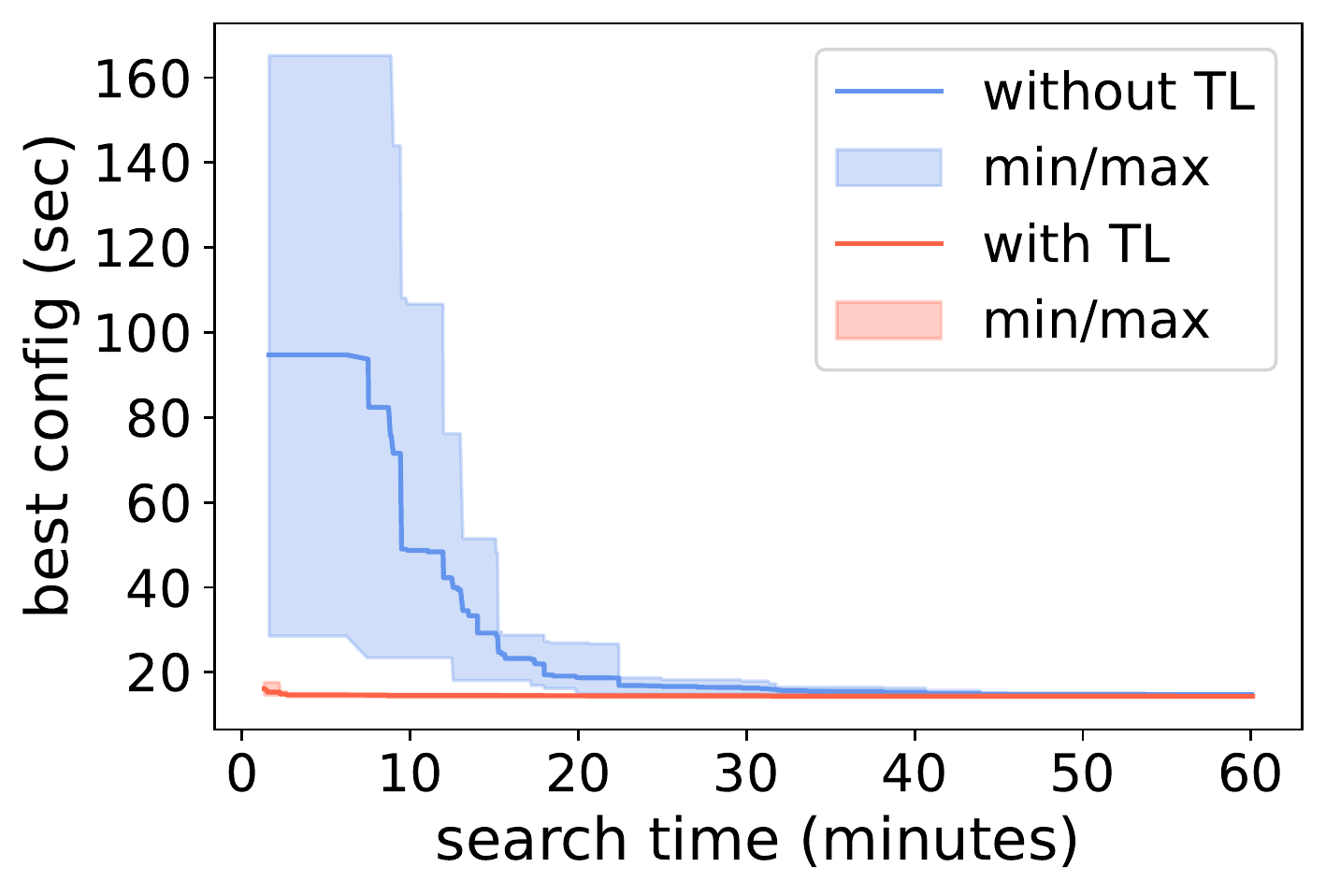}
        \caption{\revision{}{8n-2s-20p}}
        \label{fig:Experiments:tl:8-T-T}
    \end{subfigure}
    \begin{subfigure}[b]{0.32\textwidth}
        \centering
        \includegraphics[width=\textwidth]{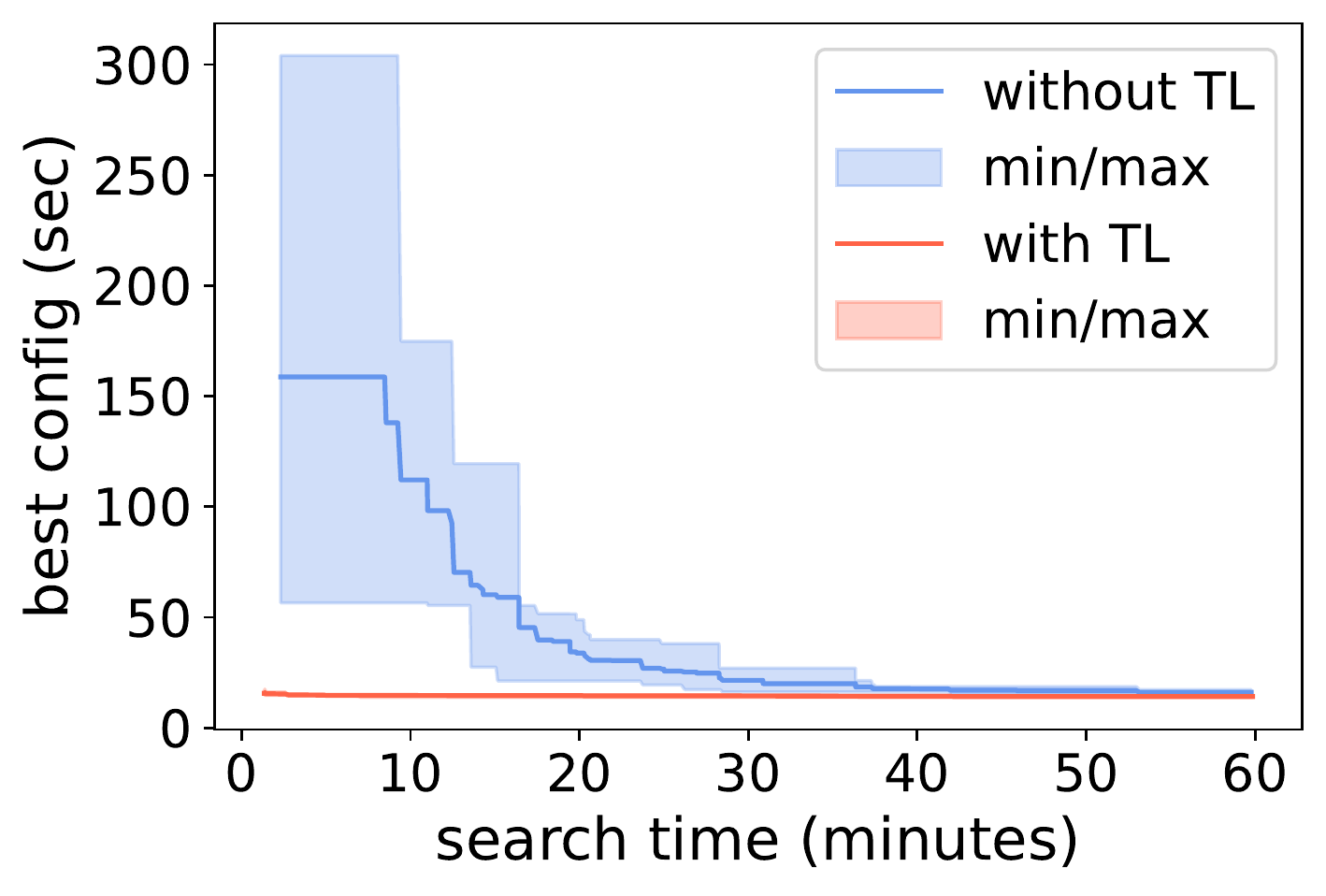}
        \caption{\revision{}{16n-2s-20p}}
        \label{fig:Experiments:tl:16-T-T}
    \end{subfigure}
    \begin{subfigure}[b]{0.32\textwidth}
        \centering
        \includegraphics[width=\textwidth]{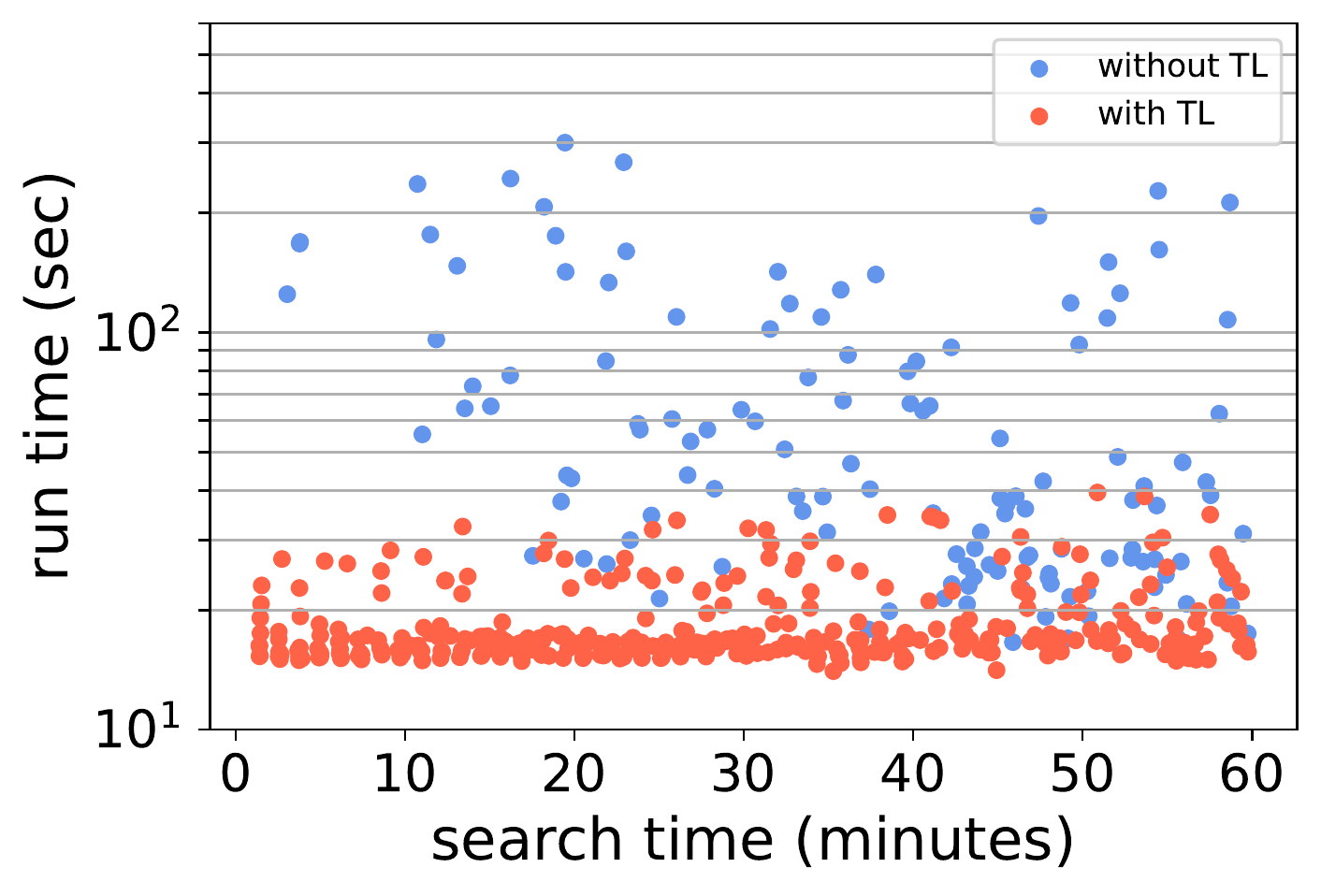}
        \caption{\revision{}{16n-2s-20p} (scatter plot)}
        \label{fig:Experiments:tl:16-T-T-scatter}
    \end{subfigure}
    
    \caption{Evolution of the run time of the best configuration found as the search time progresses (average, min, and max across 5 repetitions) for the 5 setups described in Section~\ref{sec:Evaluation:methodology:setup}, with transfer learning added for the last 4 setups.}
    \label{fig:Experiments:tl}
\end{figure*}

Figures~\ref{fig:Experiments:tl} (a) to (e) show the evolution of the best configuration found by DeepHyper with and without TL. TL enables converging toward the best configuration almost immediately, whereas not using TL leads to DeepHyper taking up to 40 minutes to converge in the worst case (\revision{}{16n-2s-20p}). Only when the parameter space and workflow change significantly (\revision{}{4n-1s-11p} to \revision{}{4n-2s-16p}) does transfer-learning take a few minutes to converge, still outperforming its non-TL counterpart.

These graphs may seem like VAE-ABO simply reuses the best configuration previously found and runs with it. This is not what happens, however, and we provide Figure~\ref{fig:Experiments:tl} (f) to demonstrate it. This figure shows all the evaluations done with and without TL (red and blue respectively) for one of the \revision{}{16n-2s-20p} runs. It shows that TL makes DeepHyper start off in the high-performing region of the parameter space and focuses on exploring this region more than a non-TL run would do. This strategy leads to lower run times per evaluation, hence more evaluations and a more refined result.



\subsection{Autotuning with different models}
\label{sec:Evaluation:all}

In the preceding section we used a random forest method in DeepHyper to model the objective function. One may wonder, however, whether a random sampling of the search space would find a high-performing configuration as well (and if so how fast), and whether another model could be used instead of random forest, in conjunction with VAE-ABO.

In this section we repeat all our previous experiments using two other modelling approaches: RAND (random sampling) and GP (Gaussian process). We chose the latter in particular because it is used in GPtune~\cite{Liu2022GPtune}, another autotuning framework for HPC, and because of its popularity in the field of Bayesian optimization.

\begin{figure*}
    \centering
    
    \begin{subfigure}[b]{0.32\textwidth}
        \centering
        \includegraphics[width=\textwidth]{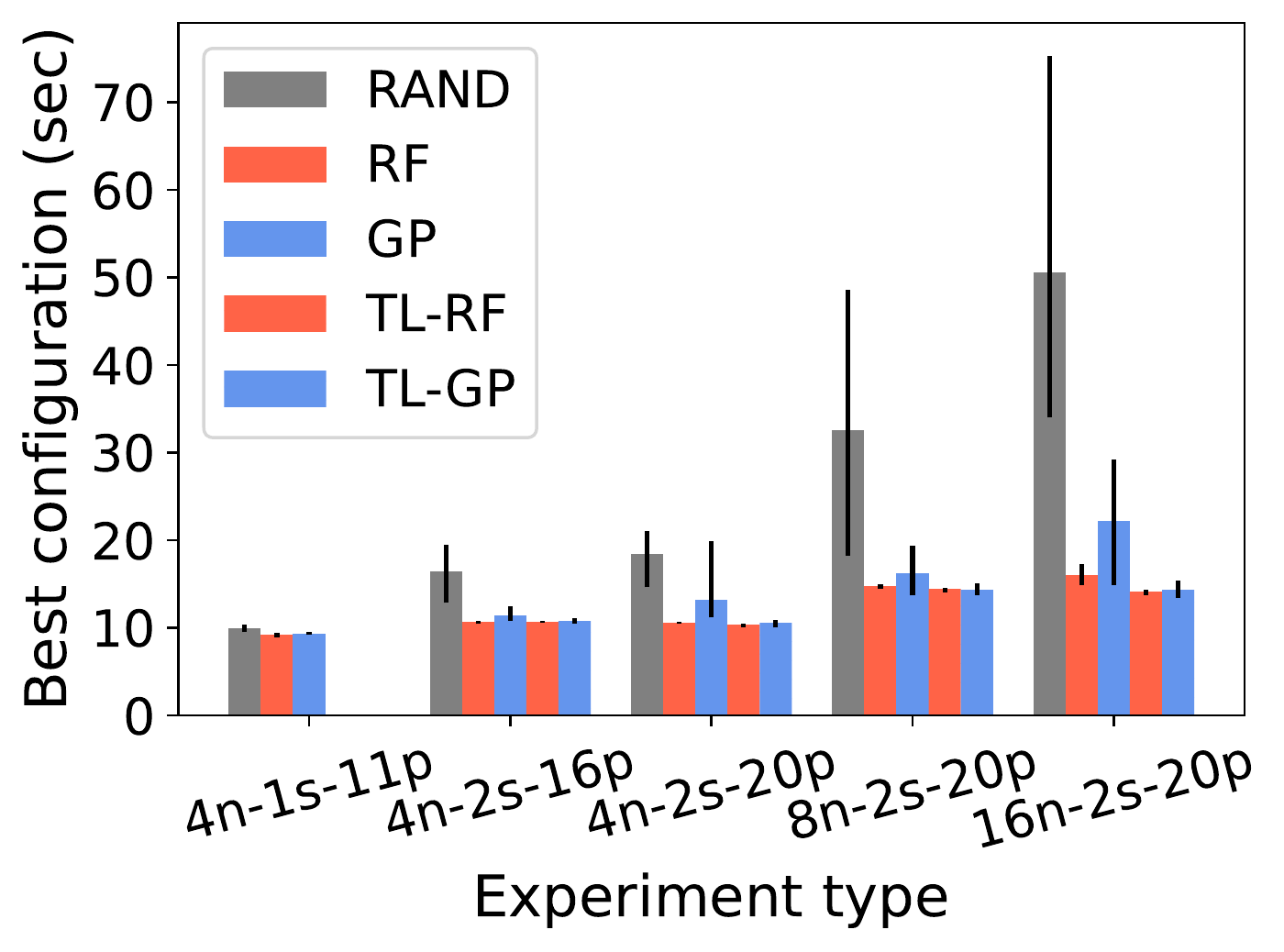}
        \caption{Best configuration overall}
        \label{fig:dh:best}
    \end{subfigure}
    \begin{subfigure}[b]{0.32\textwidth}
        \centering
        \includegraphics[width=\textwidth]{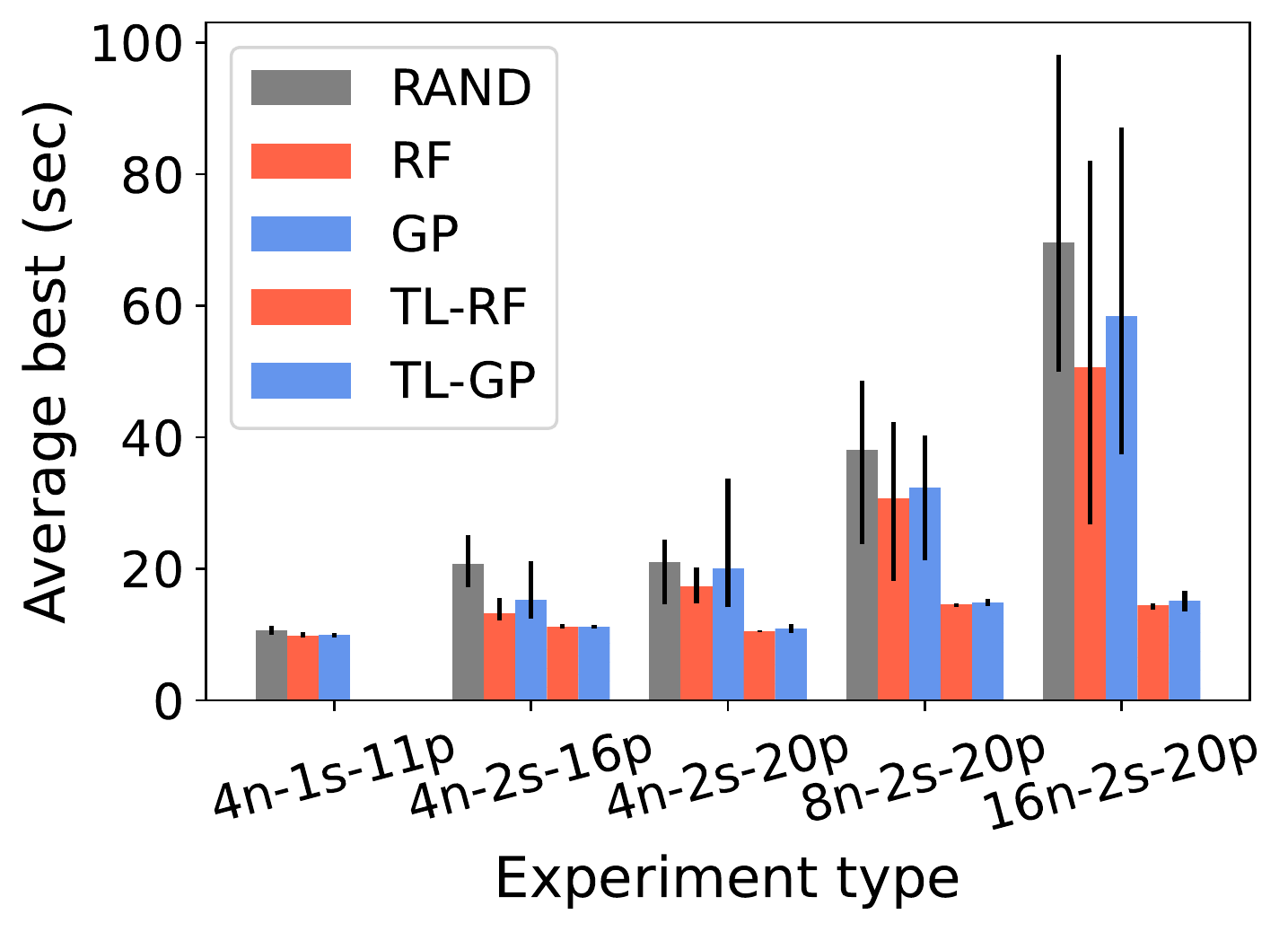}
        \caption{Mean best configuration}
        \label{fig:dh:avgbest}
    \end{subfigure}
    \begin{subfigure}[b]{0.32\textwidth}
        \centering
        \includegraphics[width=\textwidth]{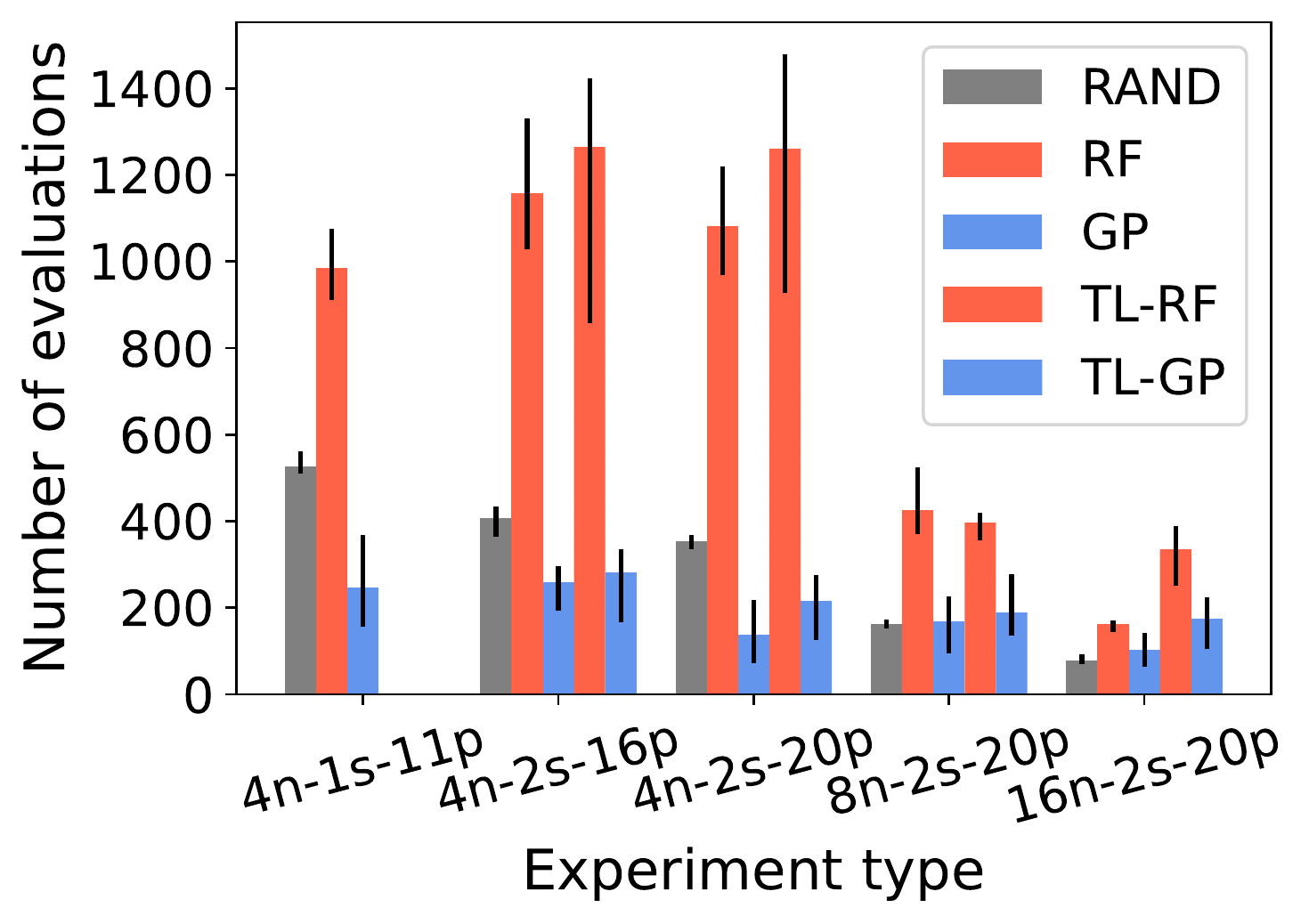}
        \caption{Number of evaluations}
        \label{fig:dh:evals}
    \end{subfigure}
    
    \begin{subfigure}[b]{0.32\textwidth}
        \centering
        \includegraphics[width=\textwidth]{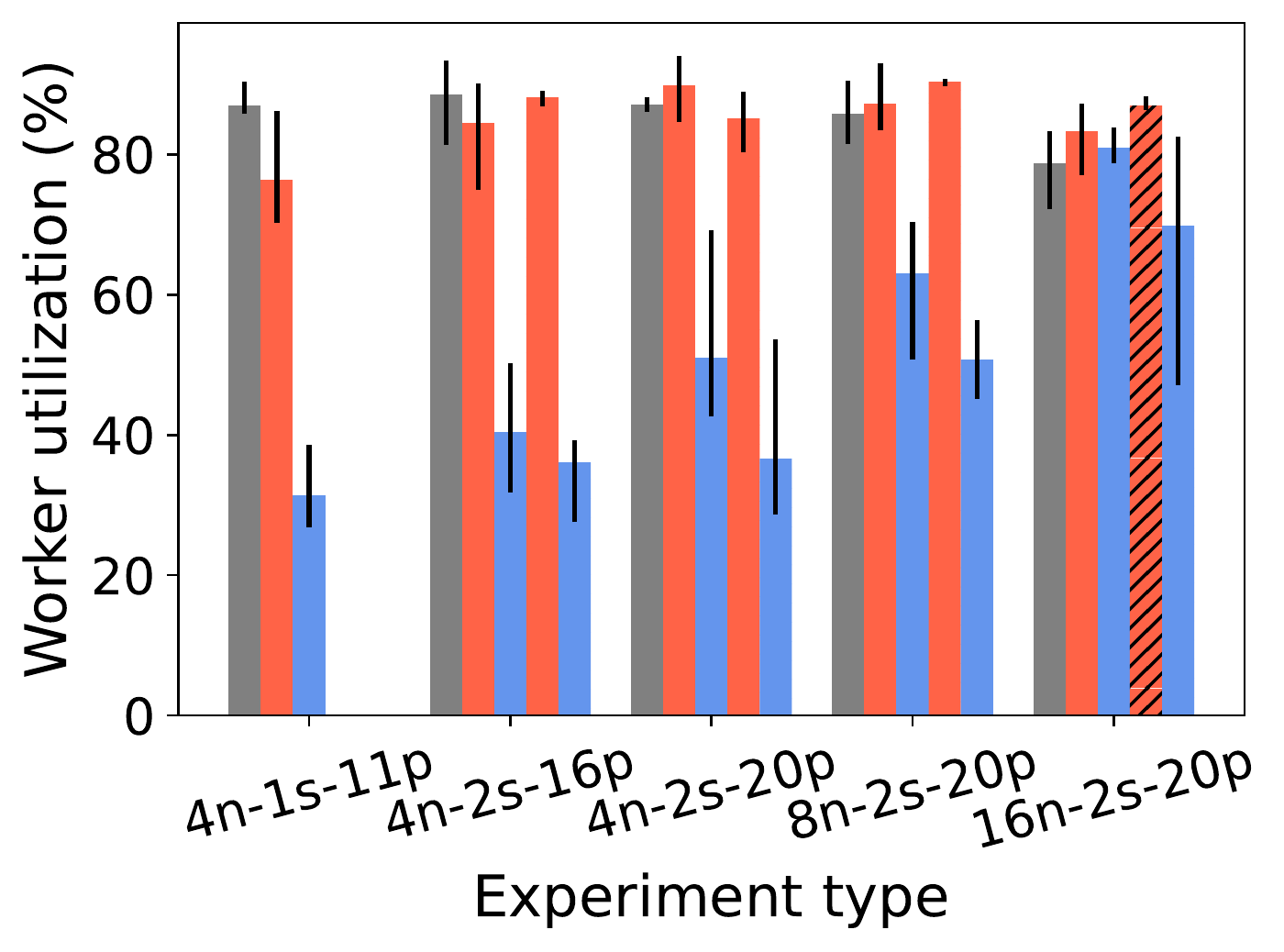}
        \caption{Worker utilization}
        \label{fig:dh:utilization}
    \end{subfigure}
    \begin{subfigure}[b]{0.32\textwidth}
        \centering
        \includegraphics[width=\textwidth]{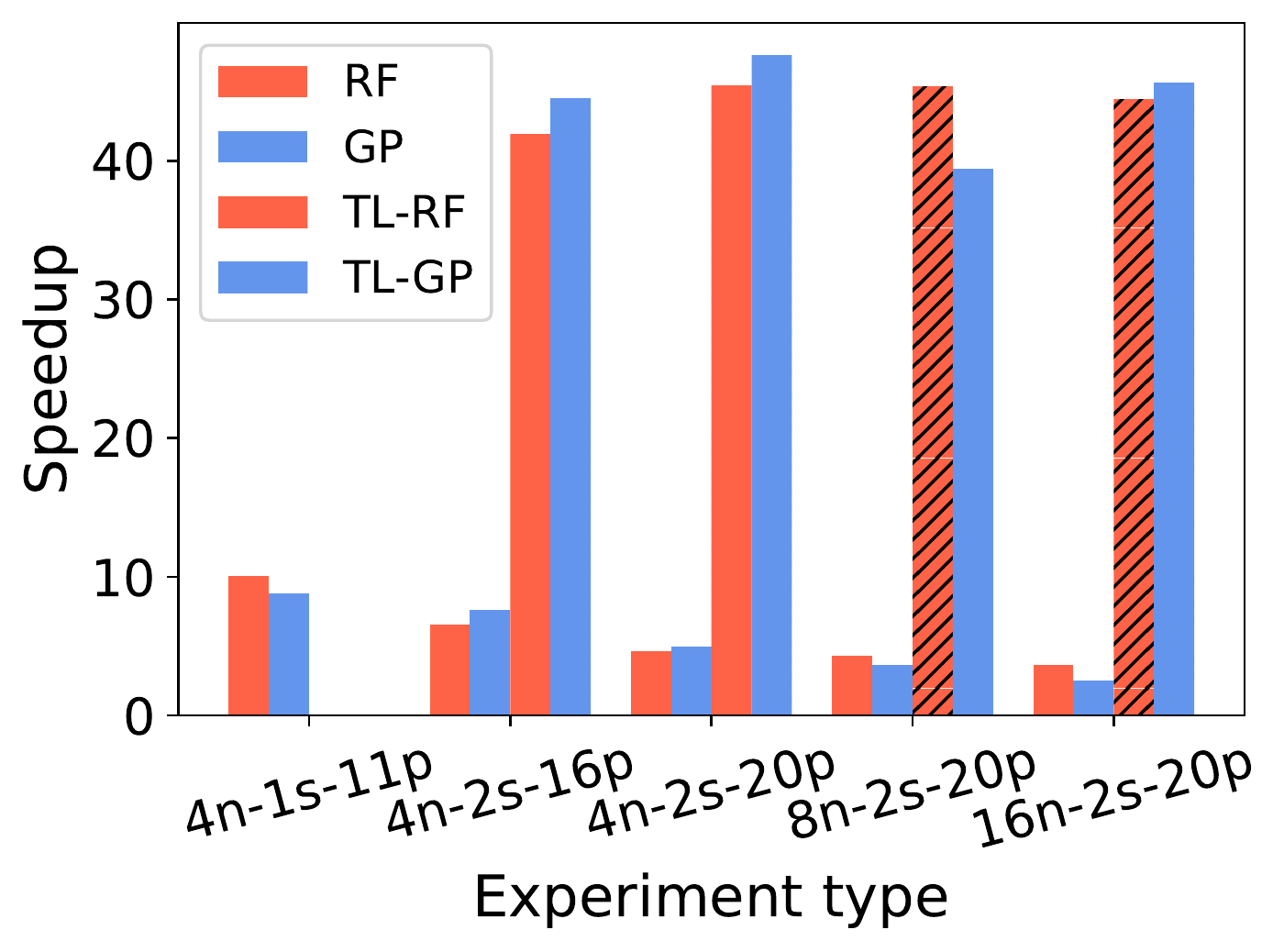}
        \caption{Speedup}
        \label{fig:dh:speedup}
    \end{subfigure}
    \begin{subfigure}[b]{0.32\textwidth}
        \centering
        \includegraphics[width=\textwidth]{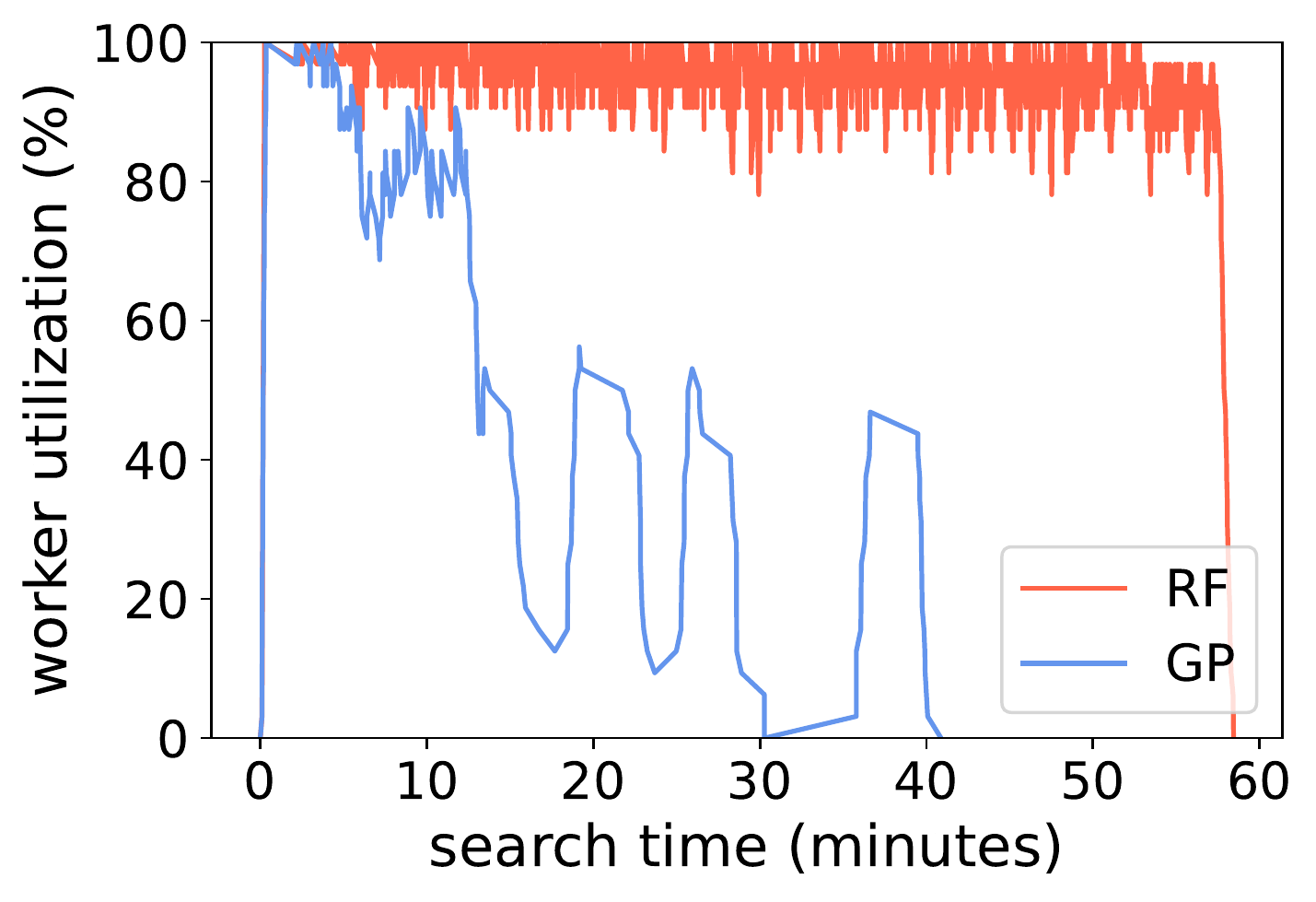}
        \caption{Example worker utilization}
        \label{fig:dh:rf_vs_gp}
    \end{subfigure}
    
    \caption{(a-e): Results of running DeepHyper with a random sampling (RAND), random forest (RF), and Gaussian process (GP) model, with and without VAE-ABO-based transfer learning. Each bar represents the average of 5 runs. Error bars represent the minimum and maximum. (f): Example worker utilization with RF and GP (one job with \revision{}{4n-2s-20p} setup, non-TL).}
    \label{fig:Experiments:dh}
\end{figure*}

Figure~\ref{fig:Experiments:dh} shows the results of these experiments in the form of the four metrics described in Section~\ref{sec:Evaluation:methodology:metrics}. Figure~\ref{fig:dh:best} shows that all models outperform a random-sampling approach in terms of the best configuration achieved. Figure~\ref{fig:dh:avgbest} showcases the benefit of transfer-learning using VAE-ABO  with both an RF model and a GP model, with both RF and GP reaching similar best run times when TL is used in all experiments. Figure~\ref{fig:dh:speedup} shows that a more than $40\times$ speedup over random search can be achieved when using transfer learning, while not using TL enables only $2.5\times$ to $10\times$ speedup.

Where RF and GP start to differ widely is in terms of number of evaluations (Figure~\ref{fig:dh:evals}) and worker utilization (Figure~\ref{fig:dh:utilization}). The GP model is much more computationally expensive to update than RF. GP has a complexity of $O(n^3)$ at every update, where $n$ is the number of evaluations performed up to the update.  Figure~\ref{fig:dh:rf_vs_gp} shows the worker utilization for two example jobs with GP and RF (\revision{}{4n-2s-20p} setup without TL). We can see that, contrary to RF, which maintains a near-100\% utilization, GP quickly reaches a point where it takes several minutes to update and decide on the next configurations to evaluate, leading to poor worker utilization.


\subsection{Comparison with state-of-the-art frameworks}
\label{sec:Evaluation:surrogate}

\begin{figure*}
    \centering
    
    \begin{subfigure}[b]{0.32\textwidth}
        \centering
        \includegraphics[width=\textwidth]{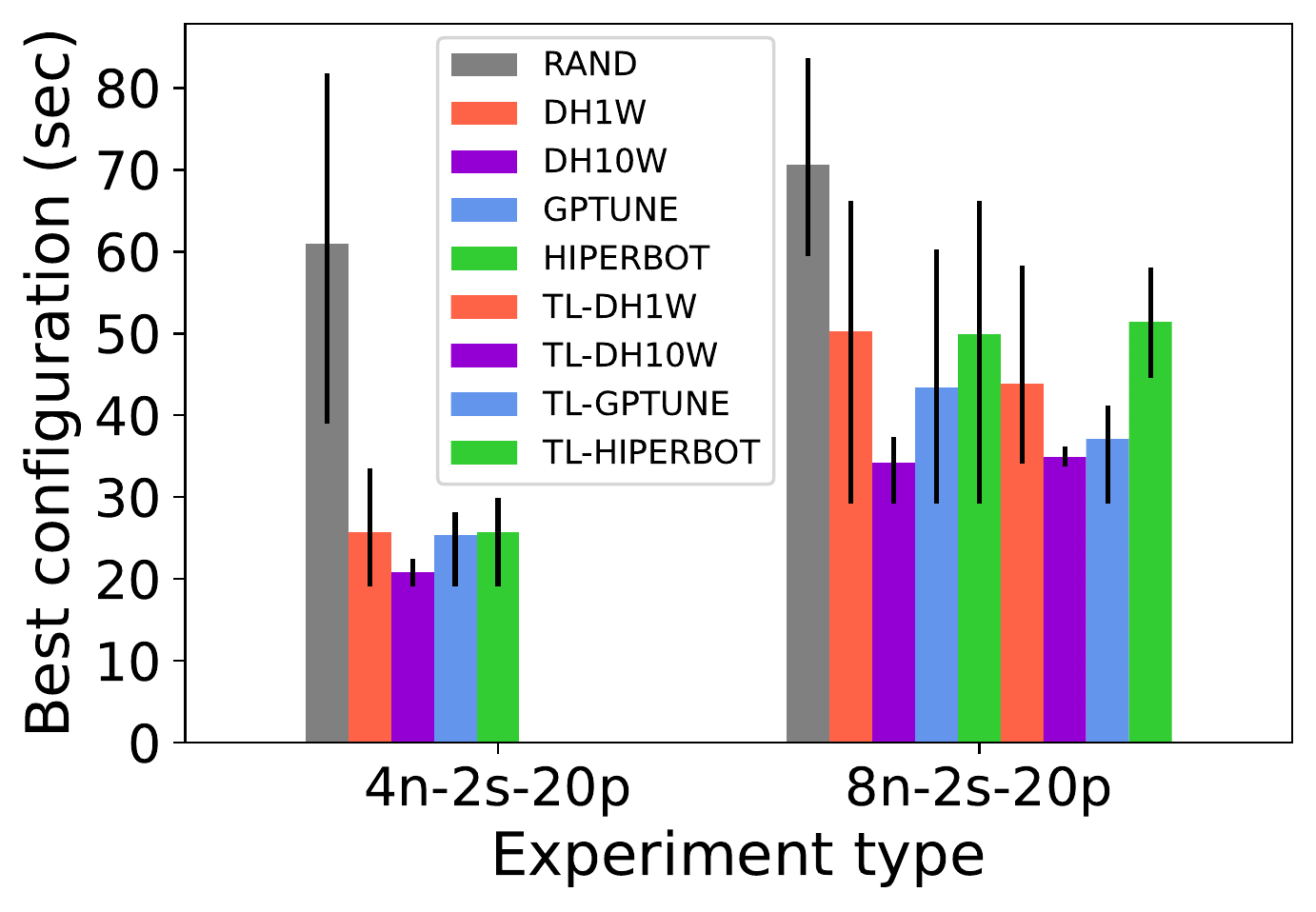}
        \caption{Best configuration overall}
        \label{fig:surrogate:best}
    \end{subfigure}
    \begin{subfigure}[b]{0.32\textwidth}
        \centering
        \includegraphics[width=\textwidth]{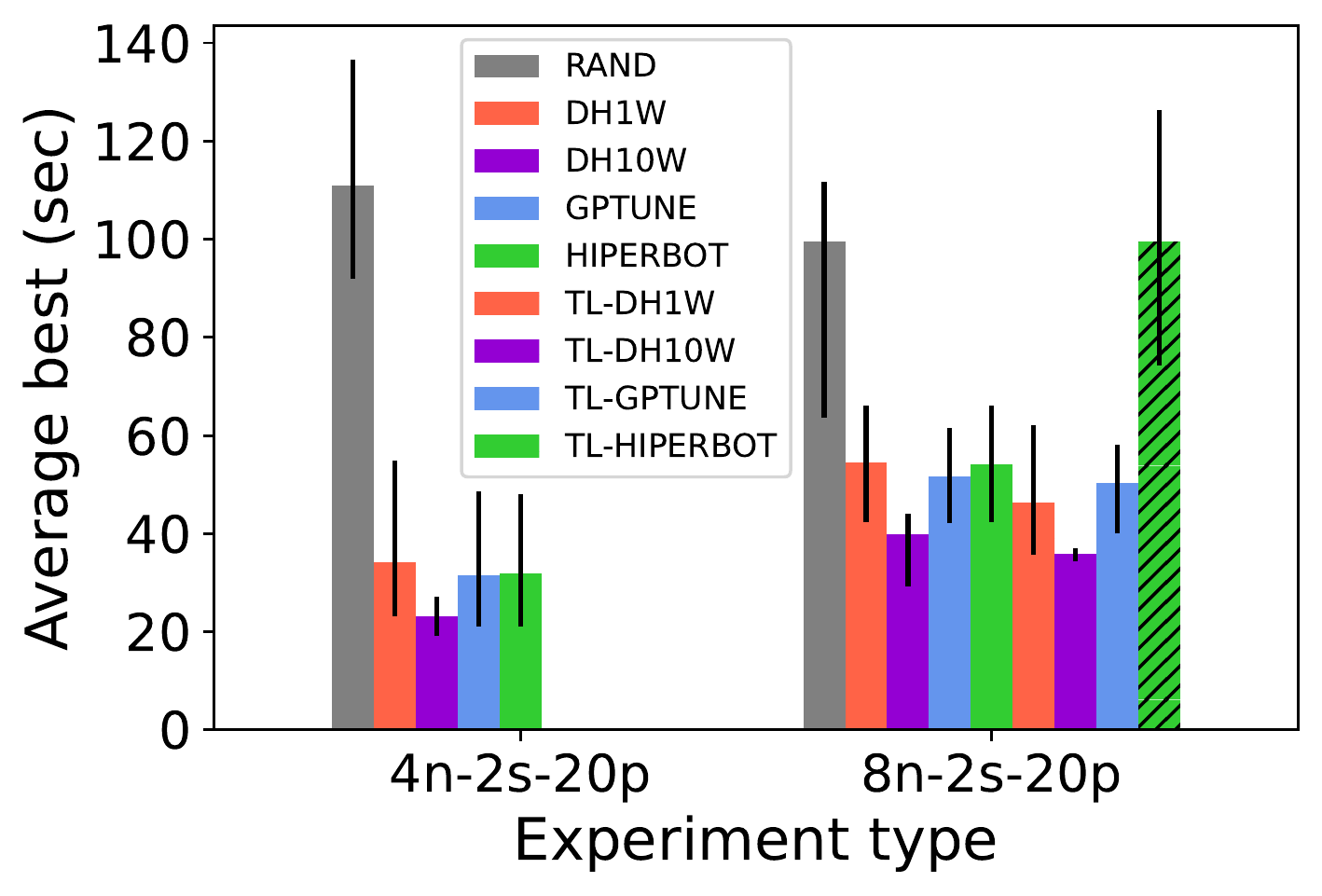}
        \caption{Mean best configuration}
        \label{fig:surrogate:avgbest}
    \end{subfigure}
    \begin{subfigure}[b]{0.32\textwidth}
        \centering
        \includegraphics[width=\textwidth]{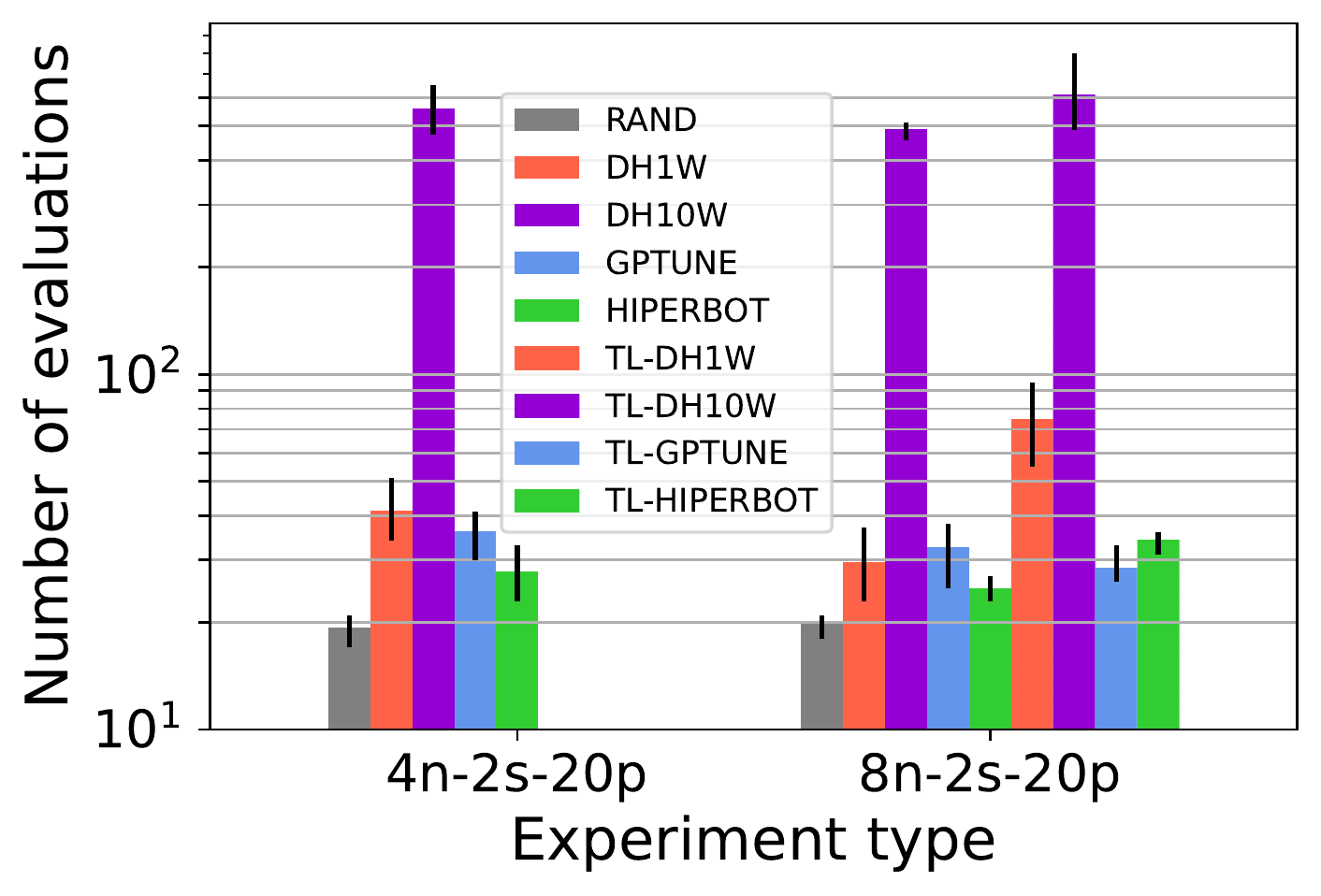}
        \caption{Number of evaluations}
        \label{fig:surrogate:evals}
    \end{subfigure}
    
    \caption{Comparison between DeepHyper, GPtune, and HiPerBOt, with the \revision{}{4n-2s-20p} and \revision{}{8n-2s-20p} setups, with and without transfer learning. Each bar represents the average of 5 runs. Error bars represent the minimum and maximum. DH1W and DH10W correspond to DeepHyper with 1 worker and 10 workers respectively.}
    \label{fig:Experiments:surrogate}
\end{figure*}

In this section, we compare our approach and implementation with two state-of-the-art parameter tuning frameworks: GPtune~\cite{Liu2022GPtune} and HiPerBOt~\cite{menon2020auto}. We selected these packages specifically because they have transfer-learning capabilities and have been used for HPC autotuning. However, we have to work around some limitations of these packages.

\textbf{GPtune.} GPtune works in two phases. During its first phase (sampling phase), it evaluates a large number of configurations using random sampling. 
In its second phase (modeling phase), it uses the result of the sampling phase to train a GP model. Technically speaking, GPtune's first phase (random search) can be parallelized, wherein the parameter configurations can be evaluated simultaneously; however, this feature was not implemented in the software at the time of our experimental campaign.

\textbf{HiPerBOt.} HiPerBOt, which relies on Bayesian optimization, does not perform any concurrent evaluation either. Note also that HiPerBOt is not an actual framework available online, so we obtained the source code from the authors and adapted it to our use case to present a fair comparison. For instance, in their paper the authors use an exhaustive enumeration of the discrete parameter space as input for transfer learning. While this is feasible for relatively small search space, it is impossible for large search space such as the one that we have. Therefore, to optimize HiPerBOt with the source data, we randomly sampled $10000$ configurations from the search space and trained on it. To perform transfer learning, we used only the configurations from HiPerBOt's optimization with the source data evaluated over a search time of one hour.


Neither of these packages is able to perform transfer learning on parameter spaces that are different. For these reasons, we performed the comparison on a single machine and only with the \revision{}{4n-2s-20p} and \revision{}{8n-2s-20p} setups, which do not require changing the parameter space. We also replaced the actual HEP workflow with a surrogate model of its performance, obtained by training a random forest regressor on the data from the preceding section's RAND runs. This surrogate model will estimate the run time for an input configuration and then sleep for this amount of time before returning it. This method makes all the experiments in this section fully reproducible on a laptop. We allow each method to run for one hour and feed them the same 10 initial random samples to make sure that none of them gets a head start simply by chance.

Figure~\ref{fig:Experiments:surrogate} shows the results. Overall, all frameworks manage to converge to comparatively well-performing configurations, with an obvious advantage for DeepHyper when using 10 workers. The average best configuration is also similar across frameworks, with the exception of HiPerBOt, which performs a lot worse when using transfer learning than when not using it. Where DeepHyper shines is in its number of evaluations, in particular when using transfer learning, even with a single worker. Although the results of random sampling (RAND) are included in Figure~\ref{fig:Experiments:surrogate}, we do not provide a comparison of the approaches in terms of speedup because random sampling performs very few evaluations, making the speedup essentially dependent on how "lucky" random sampling got with its few evaluations. We also leave out the comparison of worker utilization since this metric has little relevance in sequential runs.

While DeepHyper seems on par with GPtune in a sequential run in terms of best and average best configurations, it is worth remembering that GPtune is not able to parallelize its modeling phase, and even if it were, the Gaussian process algorithm does not scale and leads to poor worker utilization, as shown in the preceding section.

\revision{}{
Increasing the number of parameters will increase the time to find high-performing parameter configurations. This is often attributed to the curse of dimensionality, wherein the total number of configurations in the search space grows exponentially with respect to n. We can also observe this in the baseline BO without transfer learning (see Figure~\ref{fig:Experiments:tl}) for an increasing number of tunable parameters. The proposed transfer learning approach mitigates this effect by systematically leveraging the knowledge obtained from the related smaller dimensional tuning problems. 
}

Overall, and to the best of our knowledge, our approach is the only one that (1) enables asynchronous parallel evaluations even during modeling, and (2) enables transfer learning when changing the configuration space.



\section{Related work}
\label{sec:RelatedWork}

Our contributions are twofold: (1) the development and application of asynchronous Bayesian optimization for HPC storage service autotuning and (2) a new transfer learning approach for autotuning. Here, we review the related work with respect to our contributions and highlight the differences.

\subsection{Autotuning HPC storage services}

Autotuning for HPC applications in general is a vast field, with applications 
ranging from compile-time to run-time optimizations~\cite{balaprakash2018autotuning}. 
Regarding storage and I/O more specifically, \cite{behzad2013taming,behzad2013framework} 
uses a genetic algorithm used to enable auto-tuning of HPC I/O in the HDF5 stack. 
SHAMan~\cite{robert2020shaman} is another black-box auto-tuning framework for 
HPC I/O that allows selection between various heuristics, including regression-based 
surrogate modeling, simulated annealing, and genetic algorithms. 
Similarly in \cite{jia2020kill} the authors use a number of heuristics to perform 
autotuning of LSM-tree-based key/value storage systems, with a direct application 
to RocksDB. CAPES~\cite{li2017capes} adopts deep-Q learning, a deep reinforcement approach and
an alternative to genetic algorithms and Bayesian optimization, to optimize the 
performance of Lustre file systems, which is an alternative to genetic 
algorithms and Bayesian optimization. In contrast to our work, none of these works 
evaluated the potential for transfer learning in their approach.

When it comes to data-intensive HPC workflows, \cite{shu2020situ} relies on analytical 
models of workflow components alongside the workflow structure to avoid costly 
parameter-space explorations. However, this approach becomes limited when the 
number of parameters increases, and analytical models become difficult to derive. 
MapReduce workflows were also the focus of auto-tuning efforts, with works such 
as Gunther~\cite{liao2013gunther} using genetic algorithms to drive parameter 
space explorations.

In our study, we selected 20 parameters that we considered important for workflow 
performance. In practice, many more parameters could have been used. Selecting the 
few relevant ones could be done using methods such as Carver~\cite{cao2020carver}, 
which the authors applied to the tuning of local file systems.
\revision{}{In the context of autotuning databases, Kanellis et al.~\cite{kanellis2020too} showed that a small subset of the parameters drive are responsible for 99\% of the database's performance. Their pre-selection of relevant parameters is complementary to our work, and we could envision an autotuning framework that first narrows down its search to relevant parameters, before applying transfer-learning from previous searches.}

\subsection{Transfer learning in autotuning}

Only a handful of works have explored TL techniques in the context of autotuning.
\cite{valov2017transferring} proposed an approach to transfer performance prediction models 
of configurable software systems across different hardware platforms by learning a linear 
transfer model to map the relationship between their performance.
However, this TL approach is limited to tasks with the same number of parameters 
and the choice of the linear transfer model is restrictive.

\cite{jamshidi2017transfer} proposes a cost-aware TL approach that reduces 
the number of configurations sampled to obtain a performance model by reusing 
sampling points from a simulated source domain. This approach also assumes the 
same number of parameters across two tasks and relies on the availability of 
samples prior to building the performance models.

\cite{7530048} used TL to predict application performance on a target architecture 
using data collected from a different architecture. The approach employs 
model-based random search that uses a surrogate model from the previous 
optimization to prune and bias the optimization for the target architecture. 
However, this approach is sensitive to the data collected and surrogate models 
built on previous architectures.

\cite{marathe2017performance} employed autotuning on a small scale 
(in input problem size or node count) HPC applications and utilized TL 
to efficiently tune larger-scale HPC applications. They use a deep 
learning-based TL approach to transfer knowledge across performance 
prediction models. This approach also heavily relies on the availability 
of high-quality samples from the previous task and an accurate surrogate model.

\cite{sid2019multitask} employs a Gaussian process-based BO (GPTune) to 
autotune expensive exascale HPC applications. They use multi-task learning 
to autotune using all the previously available tasks and employ TL to new 
tasks by sampling selectively around the optimum from multi-task learning. 
Even though this approach overcomes the need for pre-sampled configurations 
and offline surrogate models, it is still restricted in its handling of 
integer and categorical parameters, as they are approximately transformed 
to real space. While the random sampling phase of GPTune can be parallelized, the search phase is inherently sequential, 
which limits the number of configurations that can be evaluated on a given time 
budget to converge to high-performing configurations. 


\cite{menon2020auto} proposes Bayesian optimization (BO) for autotuning parameter choices in
HPC applications (HiPerBOt). Their BO utilizes a Tree Parzen Estimator (TPE) 
(that uses a kernel density estimator 
and histograms for discrete parameters respectively) to select a new configurations 
to evaluate, instead of directly optimizing for the expected improvement as done by
GPTune. TPE tends to be faster, but in most cases less accurate. Their approach
to transfer learning is through the use of source data density as prior probability
that is weighted and combined with the target probabilities to select next best 
configuration to evaluate.

In contrast to previous transfer learning works in the context of autotuning, 
our VAE-ABO approach does not need sampled data from the previous optimization 
tasks to train a surrogate model, or transfer the surrogate model, or other 
transformations beyond the one-time cost of training a generative model on
a small subsample of high-performing configurations from
previous (source data) optimization. In addition, our approach can effectively 
handle the mixed-(real, integer, and categorical) search space and does so 
efficiently on large HPC systems. Moreover, the ability to transfer across tasks 
with different numbers of parameters is a unique capability of our transfer 
learning approach.



 \subsection{Transfer learning in optimization}
While TL in machine learning setting is a well-researched topic~\cite{zhuang2020comprehensive}, 
TL in the context of optimization has received less attention. Different strategies 
have been explored to transfer knowledge between optimization tasks. In the context 
of Bayesian optimization, the strategies employed include: 
(1) adopting kernels learned from previous optimization tasks~\cite{wang2018regret}, 
(2) changing the acquisition function to include past data or models~\cite{wistuba2018scalable}, 
(3) metalearning to identify easily adaptable configurations using previous tasks~\cite{wistuba2016two}, 
(4) a weighted combination of surrogate models from previous optimization tasks~\cite{feurer2018practical}, 
and (5) warm start current optimization with the solutions from previous optimization tasks through
search space pruning based on manually defined meta features~\cite{wistuba2015learning}, or search space 
geometries that are learned from historical data~\cite{perrone2019learning}. 
However, these methods have been studied primarily in the context of hyperparameter optimization 
for machine learning models. Recently, Souza et al.~\cite{souza2021bayesian} successfully 
demonstrated that informative priors chosen through expert knowledge help improve the 
accuracy and efficiency of Bayesian optimization. However, prior-guided Bayesian 
optimization in the context of transfer learning, in particular for mixed-integer 
search space, is largely unexplored. Moreover, none of these methods has been 
systematically explored in the context of autotuning and has not been scaled 
to take advantage of large HPC systems through parallelization.

To this end, our VAE-ABO approach can effectively transfer learn across 
optimization tasks with different parameter sizes and computational complexity 
(e.g., as defined by number of nodes, amount of computation, or I/O). 
\section{Conclusion}
\label{sec:Conclusion}

We developed a new transfer-learning-enabled asynchronous Bayesian optimization approach for tuning HPC storage services.  Our approach uses a dynamically updated, computationally cheap surrogate performance model to identify the promising regions of the search space and samples the input configurations for evaluation on the target platform. The main novelty is the ability to perform transfer learning, wherein the high performing configurations from previous \autotuning runs can be leveraged to bias the search. This is achieved by modeling joint probability distribution of the high-performing configurations using variational autoencoder  and using it for sampling configurations within asynchronous parallel Bayesian optimization. This enables BO to focus on the promising regions of the search space from the start of the autotuning search and allows transfer-learning-enabled BO Bayesian optimization to find high-quality configurations in short search time.

Applying our approach to a distributed storage service and its associated high-energy physics workflow, our results show that the use of transfer learning enables finding high-performing configurations in shorter times when (1) adding new workflow steps, (2) adding new parameters to tune, and (3) scaling up the workflow.

\revision{}{As future work, we plan to generalize our approach and provide a complete autotuning framework for Mochi-based services based on DeepHyper and our approach to transfer-learning. Such a generic framework brings the challenge of \emph{discovering} parameters from a \emph{schema} of a valid configuration file alongside a set of \emph{constraints}.}

\ifanon
\else
\section*{Acknowledgment}
We thank the authors of GPtune (in particular Yang Liu and Younghyun Cho) and HiPerBOt (in particular Harshitha Menon) for their time and valuable explanation of their respective software, and for making them available for comparison. We also thank Gail Pieper for proofreading and editing this paper.
This material is based upon work supported by the U.S. Department of Energy, Office of Science, Advanced Scientific Computing Research, under Contract DE-AC02-06CH11357.
\fi

\bibliographystyle{IEEEtran}
\bibliography{biblio}

\ifanon
\else
\fi

\end{document}